\documentclass[twocolumn]{aastex63}

\usepackage{enumitem}
\usepackage{amsmath}

\usepackage[T1]{fontenc}

\newcommand{\apec}{{\tt\string apec}}
\newcommand{\mekal}{{\tt\string mekal}}
\newcommand{\starburst}{{\sc starburst99}}
\newcommand{\xspec}{{\sc xspec}}

\newcommand{\Cunit}{$({\rm erg\,s}^{-1}) / (M_{\odot}\,{\rm yr}^{-1})^{\zeta}$}
\newcommand{\ergs}{erg~s$^{-1}$}
\newcommand{\Lsc}{$L_{\rm{SC}}$}
\newcommand{\Lx}{$L_{\rm{X}}$}
\newcommand{\mclus}{$m_{\rm{clus}}$}

\newcommand{\MSun}{$\mathrm{M}_{\odot}$}
\newcommand{\Msunyr}{$\mathrm{M}_{\odot}$~yr$^{-1}$}
\newcommand{\Rsc}{$R_{\rm{SC}}$}
\newcommand{\Rc}{$R_{\rm{C}}$}

\received{}
\revised{}
\accepted{21st January 2022}
\submitjournal{ApJ}

\shorttitle{X-ray emission from hot gas}
\shortauthors{Franeck et al.}

\begin{document}

\title{X-ray emission from star cluster winds in starburst galaxies}

\correspondingauthor{Peter Boorman}
\email{peter.boorman@asu.cas.cz}

\author{Annika Franeck}
\affiliation{Astronomical Institute of the Czech Academy of Sciences, Bo\v{c}n\'i II 1401, CZ-14100 Prague, Czech Republic}

\author[0000-0003-1848-8967]{Richard W\"unsch}
\affiliation{Astronomical Institute of the Czech Academy of Sciences, Bo\v{c}n\'i II 1401, CZ-14100 Prague, Czech Republic}

\author[0000-0002-4371-3823]{Sergio Mart\'inez-Gonz\'alez}
\affiliation{CONACYT-Instituto Nacional de Astrof\'isica, \'Optica y Electr\'onica, AP 51, 72000 Puebla, Mexico}

\author[0000-0002-9813-401X]{Ivana Orlitov\'a}
\affiliation{Astronomical Institute of the Czech Academy of Sciences, Bo\v{c}n\'i II 1401, CZ-14100 Prague, Czech Republic}

\author[0000-0001-9379-4716]{Peter Boorman}
\affiliation{Astronomical Institute of the Czech Academy of Sciences, Bo\v{c}n\'i II 1401, CZ-14100 Prague, Czech Republic}
\affiliation{Department of Physics \& Astronomy, Faculty of Physical Sciences and Engineering, University of Southampton, Southampton, SO17 1BJ, UK}

\author[0000-0003-2931-0742]{Ji\v{r}\'{i} Svoboda}
\affiliation{Astronomical Institute of the Czech Academy of Sciences, Bo\v{c}n\'i II 1401, CZ-14100 Prague, Czech Republic}

\author[0000-0001-6473-7085]{Dorottya Sz\'ecsi}
\affiliation{I. Physikalisches Institut, Universit\"at zu K\"oln, Z\"ulpicher Strasse 77, D-50937 K\"oln, Germany}
\affiliation{Institute of Astronomy, Faculty of Physics, Astronomy and Informatics, Nicolaus Copernicus University, Grudzi\k{a}dzka 5, 87-100 Toru\'{n}, Poland}

\author[0000-0001-7306-5698]{Vanesa Douna}
\affiliation{Instituto de Astronom\'ia y F\'isica del Espacio, CONICET-UBA, Ciudad Aut\'onoma de Buenos Aires, Argentina}

\begin{abstract}

Inspired by the excess soft X-ray emission recently detected in Green Pea galaxies, we model the soft X-ray emission (0.5 -- 2.0~keV) of hot gas from star cluster winds. By combining individual star clusters, we estimate the soft X-ray emission expected from the typically unresolved diffuse hot gas in starburst galaxies, devoid of competing emission from e.g., AGN or other unresolved point sources. We use stellar models of sub-solar metallicities ($0.02$~$Z_{\odot}$ and $0.4$~$Z_{\odot}$), and take into account supernova explosions for massive stars. For lower metallicities, we find that stellar winds do not contribute significantly ($\lesssim 3$\% of the mechanical energy) to the observed soft X-ray emission of normal star forming galaxies. For higher metallicities and possibly also for larger proportions of massive star clusters in the simulated starburst galaxies, we reproduce well the observed correlation between star formation rate and X-ray luminosity previously reported in the literature. However, we find that no combination of model assumptions is capable of reproducing the substantial soft X-ray emission observed from Green Pea galaxies, indicating that other emission mechanisms (i.e. unusually large quantities of High-/Low-Mass X-ray Binaries, Ultra-Luminous X-ray sources, a modified initial mass function, Intermediate-Mass Black Holes, or AGN) are more likely to be responsible for the X-ray excess.
\end{abstract}


\section{Introduction} 
\label{sec.intro}

The hot gas within star forming regions is often invoked to contribute to the X-ray energy budget in star-forming galaxies, yet only a limited number of detailed studies of its contribution are available \citep[see e.g.][]{Silich.etal.2005, Lopez.etal.2014, Rosen.etal.2014}. More widely-studied sources of X-ray emission in galaxies are believed to arise from X-ray coronae found in accreting supermassive black holes (aka Active Galactic Nuclei; AGN) and X-Ray Binaries. Here, X-ray Binaries refer to any compact object (typically a stellar mass black hole or neutron star) accreting material from a High-Mass or Low-Mass companion (aka High-Mass X-ray Binaries \& Low-Mass X-ray Binaries, respectively).

The relative proportions of each X-ray-emitting process to the total observed X-ray luminosity depends on specific conditions within the galaxy. In the nearby universe, the X-ray emission from individual starburst galaxies was studied by e.g., \citet{Grimes2007,Oti-Floranes2012,Oti-Floranes2014,BasuZych.etal.2013.I,Basu-Zych2016}, who attributed all the unresolved X-ray emission to the hot gas. However, corrections were not applied for unresolved compact sources. \citet{Mineo.etal.2012.I, Mineo.etal.2012.II} systematically studied the X-ray spectra of a sample of nearby star-forming galaxies. To evaluate the X-ray contribution from the hot gas, the resolved High-Mass X-ray Binary contributions to the X-rays were removed, alongside an estimated unresolved High-Mass X-ray Binary component. Contributions from other unresolved X-ray sources (including Low-Mass X-ray Binaries, cataclysmic variables, AB stars, young stellar objects \& supernova remnants) were estimated and found to be relatively weak in the soft 0.5--2~keV band. \citet{Mineo.etal.2012.II} then found the hot gas to contribute at a level of $<30\%$ to the total observed X-ray emission, and the diffuse emission in the \mbox{0.5 -- 2\,keV} band to correlate with the star formation rate (SFR) of the galaxy. 

As with the X-ray emission from hot gas, the emission from High-Mass X-ray Binaries also scales with SFR due to the rapid stellar evolution of higher-mass stars. High-Mass X-ray Binaries are hence typically short-lived and thus the contribution to the X-ray budget is broadly proportional to SFR with a modification due to metallicity. In contrast, the X-ray contribution from longer-lived Low-Mass X-ray Binaries scales with the mass of the galaxy. Empirical laws can then be derived which connect SFR to the observed X-ray luminosity of a given galaxy \citep{Ranalli2003,Mineo.etal.2012.I,BasuZych.etal.2013.I,Douna.etal.2015,Brorby.etal.2016}.

Unusually bright X-ray emission was recently reported from a class of intermediate redshift ($z\sim0.2 - 0.3$) compact starburst galaxies called ``Green Peas'' \citep{Svoboda.etal.2019}. Green Pea galaxies have attracted attention for their similarity to high-redshift galaxies near the Epoch of Reionisation -- namely low mass, low metallicity, vigorous star formation \citep{Cardamone.etal.2009}, high ionization, and high fractions of escaping ionizing ultraviolet radiation \citep[e.g.][]{Izotov2016,Izotov2018b,Kim20}. \citeauthor{Svoboda.etal.2019} analysed \textit{XMM-Newton} observations of three Green Pea galaxies, finding two to have X-ray luminosities a factor $\sim$4--6 times brighter than predicted from empirical relations between the X-ray luminosity, SFR and metallicity for star-forming galaxies. If the X-ray excess is connected to starburst activity, possible explanations include: High-Mass X-ray Binaries, Low-Mass X-ray Binaries, high numbers of Ultra-Luminous X-ray sources, modified initial mass functions, intermediate-mass black holes, or a substantial contribution from hot gas. The remaining intriguing possibility is an X-ray flux contribution from a hidden AGN. \citet{Kawamuro.etal.2019} performed relatively short exposure hard X-ray observations of the two \textit{XMM-Newton}-detected Green Pea galaxies with \textit{NuSTAR}, finding both to be non-detections $\geq$\,10\,keV. Using an established mid-infrared vs. X-ray correlation for local AGN-hosting Seyfert galaxies, \citeauthor{Kawamuro.etal.2019} showed that if this correlation holds for Green Pea galaxies, and the observed AGN-like mid-infrared colours were powered by an AGN, the central engine would have to have column densities in excess of $N_{\rm H}$\,$\gtrsim$\,2$\times10^{24}$\,cm$^{-2}$ to explain the \textit{NuSTAR} non-detections. Thus, none of the hypotheses have been completely satisfactory to-date to explain the power source responsible for Green Pea X-ray emission and none can currently be directly proven observationally.

The contribution of X-ray emission from hot diffuse gas in star-forming regions has been analysed and considered in a multitude of ways in previous theoretical studies. \citet{Cervino.etal.2002} computed the X-ray emission from diffuse gas in a single stellar population heated by stellar winds and supernovae. An efficiency parameter was used to represent the fraction of output mechanical energy that heats the gas. However, the efficiency of this process is very difficult to determine and many different values have been used in the literature. 
\citet{Stevens.Hartwell.2003} modeled the properties of a stellar cluster based on the analytical description of the wind by \citet{Chevalier.Clegg.1985}. The authors compared model radial temperature profiles and diffuse emission of the gas to a sample of nearby clusters observed by \emph{Chandra}. Once again, disentangling the contribution of the diffuse gas from that of other sources was not straightforward. 
Aiming to overcome this problem, \citet{Oskinova2005} followed a similar approach but additionally included the star clusters' evolution and emission of stars, before comparing the predictions to an observed sample of clusters in the Large Magellanic Cloud (LMC). The author attributed the differences between the observed and predicted diffuse X-ray emission to the evolutionary stages of the clusters and to the evolution of X-ray-emitting low-mass stars.

\citet{Marcolini2005} carried out a series of 2-D hydrodynamical simulations of pc-scale conductive clouds immersed in a galactic-scale wind. The authors found that as the clouds are shock-processed, the resulting soft X-ray emission is dominated by cloud material even if the wind metallicity is several times higher than that of the clouds.

To gain insight into the X-ray emission contribution from hot gas in galaxies beyond that provided by empirical relations (as used in \citeauthor{Svoboda.etal.2019}), here we use a hydrodynamical approach that follows the thermalization of the kinetic energy deposited by stellar winds and supernova explosions within stellar clusters \citep{Silich.etal.2004, Silich.etal.2011, Wunsch.etal.2007, Wunsch.etal.2011}. For a pre-defined population of stars in a given volume, our code -- WINDCALC \citep{Wunsch.etal.2017} -- computes the radial profiles of gas density and temperature, while accounting for wind collisions from individual stars as well as the mass and energy deposited by supernova explosions. The code has been tested against other numerical codes and observations, yielding robust results \citep{Silich.etal.2005, Wunsch.etal.2007, Wunsch.etal.2011}. Here we incorporate the X-ray emission from such populations of stars. Our work is originally driven by the unexplained X-ray luminosity of Green Pea galaxies, but is equally useful for understanding gas emission in less extreme objects such as ``ordinary'' star-forming galaxies as well as individual star clusters.

The paper is structured as follows: the numerical code that defines the set-up of the star clusters and galaxies, as well as the computation of X-ray luminosities is introduced in Section~\ref{sec.simulation}. The results of the modeling, the role of various parameters, and the comparison with observational data are presented in Section~\ref{sec.results}, and discussed in Section~\ref{sec.discussion}. 
Finally, the summary and conclusions are provided in Section~\ref{sec.summary}.

\section{Numerical simulations of star clusters, X-ray emission, and galaxies}
\label{sec.simulation}

We aim to calculate the X-ray emission from shocked stellar winds in a population of star clusters within starburst galaxies, such as those possible in Green Pea galaxies. First we calculate the expected X-ray emission from individual star clusters for a grid of cluster masses and cluster ages. We then use those star clusters to construct a starburst galaxy according to the SFR of the galaxy. Once the star cluster population has established mass and age distributions, we integrate the X-ray spectra of the individual star clusters to give the X-ray spectrum of the starburst galaxy. From the resulting spectrum we calculate the X-ray luminosity \Lx\ within the soft \mbox{0.5 -- 2.0~keV} energy band. 
We then estimate \Lx\ for starburst galaxies as a function of SFR, and investigate how our results change with different assumptions.

\subsection{Simulation of a star cluster} 
\label{sec.starcluster}

\subsubsection{Distribution of stars within a star cluster}
\label{sec.starcluster.diststars}
We follow the semi-analytic approach of \citet{Wunsch.etal.2017} to model the hydrodynamics of the shocked stellar winds in individual star clusters, with masses ranging from $10^3$ to $10^8$~$M_{\odot}$. For each cluster, we draw a distribution of stars according to the standard initial mass function with an index of $\alpha = -2.3$ at the highest mass end \citep{Kroupa2001}, and with star masses $m_{\rm{star}}$ between 0.01 and 120~$M_{\odot}$. The stars are created instantaneously, and are distributed according to a Schuster profile of $\rho_{\rm{star}} \propto [1 + (r/R_{\rm{C}})^2]^{-b}$ where $R_{\rm{C}}$ is the core radius and $b = 3/2$ \citep{Palous.etal.2013, Tenorio-Tagle.etal.2015}. The distance from the centre is denoted by~$r$. Furthermore, we use the cutoff radius \Rsc\ to define the radius of the star cluster. If not indicated otherwise, all clusters have a core radius of $R_{\rm{C}} = 1$~pc and a cutoff radius of $R_{\rm{SC}} = 3$~pc.

\subsubsection{Stellar evolution models}
\label{sec.starcluster.evolution}

We follow the evolution of individual stars with stellar models taken from the BoOST project \citep{Szecsi.etal.2020}. These models were computed with the `Bonn' Code \citep[e.g.][and references therein]{Szecsi.etal.2015} assuming two different metallicities $Z = 0.02\,Z_{\odot}$ as for the dwarf galaxy IZw18, and $Z = 0.4\,Z_{\odot}$ as for the Large Magellanic Cloud (LMC). We consider low metallicities for consistency with the measured sub-solar values for Green Pea galaxies (e.g., \citealt{Izotov.etal.2011}). Additionally, in Section~\ref{sec.results_CHE} another set of stellar models from the BoOST project are tested, which incorporate chemically-homogeneous evolution. We interpolate linearly between the stellar models using the code \textsc{synStars} \citep[e.g. sect.~4 of][]{Szecsi.etal.2020} to get synthetic populations that consist of smoothly interpolated stellar tracks. Additionally, our treatment of massive stars exploding as supernovae is similar to the procedure used in \starburst\ \citep{Leitherer.etal.1999}. 
In low metallicity environments, (e.g., for the $Z = 0.02\,Z_{\odot}$ case), stars with masses between 8 and 40~$M_{\odot}$ explode as regular supernovae (\citealt{Heger.etal.2003}, also cf. the detailed discussion in Section~3.5 of \citealt{Szecsi.Wunsch.2019} about this question), while for the higher LMC-like metallicity we assume that all stars with masses $m_{\rm{star}} \geq 8\,M_ {\odot}$ explode. This results in different starting times of the supernovae in the framework of the cluster's lifetime. Since massive stars live for shorter periods of time, the supernovae in the higher-metallicity environment ($Z = 0.4\,Z_{\odot}$) arise at $t_{\rm{clus}} \sim 2$~Myr. In the lower metallicity environment ($Z = 0.02\,Z_{\odot}$) supernovae start from $t_{\rm{clus}} \sim 4.5$~Myr. Each supernova contributes an energy of $10^{51}$~erg to the gas and returns a mass of $m_{\rm{SN}} = m_{\rm{star}} - m_{\rm{remnant}}$ to the free gas available in the cluster, where the remnant mass is fixed to $m_{\rm{remnant}} = 1.4$~$M_{\odot}$.\footnote{We note that this assumption is unlikely to hold for the highest-mass supernovae in our simulations, but the predicted difference in observed X-ray emission from the resulting clusters is expected to be small. See Sz\'{e}csi et al., in prep. for more details.} 

We distribute the supernovae in a way that the supernova rate is constant until the end of each stellar track, but changing to the next stellar track results in a different supernova rate. Thus the supernova rate is in general variable throughout a full simulation. 

At the state of the art one cannot predict in which type of supernova a star will explode based on its stellar evolution track, and such considerations are outside the scope of this work. Therefore, predictions about chemical changes due to supernovae cannot be well constrained. Thus we restrict ourselves to use the final surface chemical composition of each star for calculating X-ray spectra.

\subsubsection{Radial profiles of the star clusters}
\begin{figure*}
	\includegraphics[width = \textwidth]{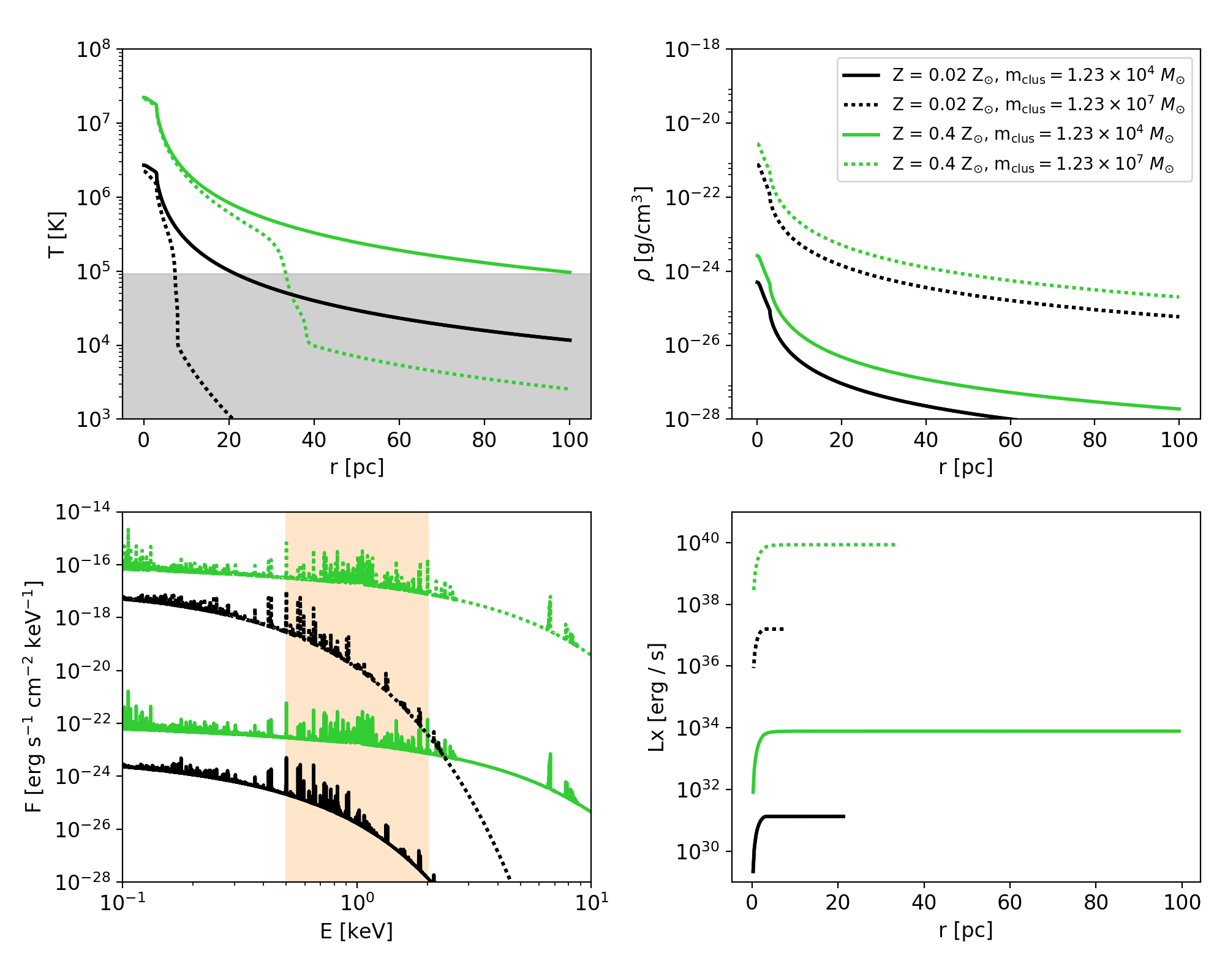}
	\caption{Radial profiles of the gas temperature (\textit{upper left}), and gas density (\textit{upper right}) for clusters of two different cluster masses (solid and dashed lines, respectively) and metallicities ($0.02$~$Z_{\odot}$ and $0.4$~$Z_{\odot}$ in black and green, respectively). The profiles are shown for a cluster evolutionary time of $t = 1.5$~Myr. In the \textit{lower left panel} we present the corresponding X-ray spectra by integrating up to an aperture radius $R_\mathrm{ap} = 100$\,pc. Higher densities yield brighter X-ray fluxes whilst line emission is more pronounced at higher metallicities. The orange shaded area marks the soft X-ray energy band (0.5 -- 2.0~keV), used in this study. The \textit{lower right panel} shows the cumulative luminosity of this energy band from the \xspec\ components as a function of the aperture $R_\mathrm{ap}$ of the star cluster. For all cases, the dense, hot gas in the centre of the cluster dominates the X-ray emission of the whole cluster. The grey shading in the upper left panel indicates temperatures that do not contribute to the X-ray emission in the \apec\ model. Hence, the different curve lengths in the lower right panel -- see the text for details.}
	\label{fig.Profiles}
\end{figure*}
We follow the evolution of each star cluster up to an age of $t_{\rm{clus, max}} = 20$~Myr. Our model provides the energy loss, mass loss and chemical composition of the gas as a function of time.

Assuming spherical symmetry, we calculate the radial profiles of the gas temperature and gas density in the star cluster up to a distance of $R_{\rm{max}} = 100$~pc from the cluster centre. Figure~\ref{fig.Profiles} shows examples of the radial profiles for two different cluster masses and the two metallicities at a cluster's evolutionary time of $t_{\rm{clus}} = 1.5$~Myr, before the onset of supernovae. We find the temperatures in the cluster's centre to depend mainly on the gas metallicity. At larger radii, the temperature in the higher mass star cluster drops faster compared to the more massive cluster, due to the high mass cluster being in a strongly radiative regime \citep[see][for further details]{Wunsch.etal.2007, Silich.etal.2004}. 

Figure~\ref{fig.Profiles} also shows the gas density profiles of the clusters. The mass of the cluster mainly determines the density of the gas due to the fixed cutoff radius \Rsc -- i.e. while the volume of the clusters remains fixed, the amount of stars grows proportionally with \mclus. Thus the contribution to the stellar wind increases and the gas density scales linearly with \mclus. Further, we find the gas density of two clusters with the same mass to be larger for higher metallicities due to the higher mass loss of stars at higher metallicity.

\subsection{X-ray spectrum of a star cluster}
\label{sec.Xray_Starcluster}

\subsubsection{Calculation of the X-ray emission with \xspec}
Based on the temperature and density profiles of the star clusters (see top panels in Figure~\ref{fig.Profiles}), we calculate the X-ray spectra by integrating up to $R_\mathrm{ap} = 100$~pc. We use the model \apec\ \citep[Astrophysical Plasma Emission Code, v.12.10.1][]{Smith01apec} within the \xspec\ package \citep{Arnaud.1996} to calculate the continuum and line emission from optically thin collisionally-ionized gas.
Assuming spherical symmetry, for each cluster, we take into account all gas within the aperture, i.e. $r < R_\mathrm{ap}$ (this neglects the contribution from the gas at large radii projected into the aperture, however, in Section~\ref{sec.Xray_Spectra.luminosity} it is shown that this contribution is very small). We divide the sphere into 300 shells distributed equidistantly in the logarithm of temperature. Each of the 300 shells is considered as a single \apec\ component, whose X-ray contribution is calculated based on four input parameters:
\begin{itemize}[noitemsep]
	\item[\textit{(i)}] the gas plasma temperature $k_BT$, where we use the mass weighted temperature of each shell, and the Boltzmann constant $k_B$,
	\item[\textit{(ii)}] the chemical abundances derived from the stellar models (see Section~\ref{sec.starcluster.evolution}),
	\item[\textit{(iii)}] the redshift, which we set to $z = 0$ to obtain the spectrum in the intrinsic rest frame.

	\item[\textit{(iv)}] a normalization parameter, which is the emission measure scaled by the distance, as $10^{-14} / \{4 \pi [D_{\rm{A}}\,(1+z)^2] \} \times \int n_{\rm{e}} n_{\rm{H}} \rm{dV}$. Here, $D_{\rm{A}}\,(1+z)^2$ corresponds to the luminosity distance $D_{\rm{L}}$ which we set to the value for SDSS\,J074936.77+333716.3 (aka \lq GP1\rq); $D_{\rm{L}} = 1437.9$~Mpc \citep{Svoboda.etal.2019}\footnote{The distance was calculated from redshift $z$\,=\,0.2733 with cosmological parameters: $H_{0}$\,=\,67.8\,km\,s$^{-1}$\,Mpc$^{-1}$, $\Omega_{\rm M}$\,=\,0.308, $\Lambda_{0}$\,=\,0.69.}. Despite step (iii), we applied a representative Green Pea distance to the normalization to obtain appropriate flux levels that could be directly compared to real measurements from Green Peas \citep{Svoboda.etal.2019}. The number densities of electrons and hydrogen ($ n_{\rm{e}}$, $ n_{\rm{H}}$, respectively) result from the stellar models (see Section~\ref{sec.starcluster.evolution}).
\end{itemize}
Finally, we sum the X-ray emission of the individual shells to obtain the X-ray emission of the whole star cluster.

\subsubsection{X-ray spectra and luminosity of the star clusters}
\label{sec.Xray_Spectra.luminosity}
For the modelled star clusters introduced in Section~\ref{sec.starcluster}, we show in the lower left panel of Figure~\ref{fig.Profiles} the corresponding X-ray spectra. 
The spectra are shown in the rest frame. Only, the flux level corresponds to a source that would be located at $z \approx 0.3$, typical for Green Pea galaxies.
In general, the X-ray flux in high mass clusters is larger compared to that of the lower mass clusters due to the higher densities that ensure more mass is available for radiating X-ray emission. The temperature of the gas -- determined by the gas metallicity -- spreads the continuum to higher energies. Therefore, the spectra with $Z = 0.4\,Z_{\odot}$ drop at higher energies compared to the lower metallicity cases (green vs. black lines in Figure~\ref{fig.Profiles}). 

To quantify the X-ray emission from the hot gas, we calculate the X-ray luminosity in the soft X-ray band of \mbox{0.5 -- 2.0~keV} (shaded area in the lower left panel of Figure~\ref{fig.Profiles}) by
\begin{equation}\label{eq1}
    L_{\rm{X}} = 4\pi d^2 \int_{0.5\,\rm{keV}}^{2.0\,\rm{keV}} F(E) \rm{d}E,
\end{equation}
with the distance $d$ set to the same value for GP1 and the photon energy, $E$\,=\,$h\nu$ where $h$ is the Planck constant and $\nu$ is the photon frequency. In Figure~\ref{fig.Profiles} we show the cumulative luminosity \Lx\ as a function of the aperture. We find the inner part of the cluster to dominate the X-ray emission for all cluster masses and metallicities. The \apec\ model only includes contributions with plasma temperatures $k_{\rm{B}}T \gtrsim 8.0 \times 10^{-3}$~keV, (corresponding to $T \sim 9.2 \times 10^4$~K). Therefore, the lines in the Figure end once the temperature is beyond this value, and the temperatures not considered for the X-ray emission calculation are shaded in grey in the temperature profiles in Figure~\ref{fig.Profiles}. Further, we use an aperture radius of $R_\mathrm{ap} = 100$\;pc to provide us with a reasonable safety factor that ensures we accounted for the majority of the X-ray emission.

\subsubsection{Evolution of \Lx\ and the energy insertion rate of the star clusters}
\label{sec.Xray_Starcluster.LscLx}
\begin{figure*}
	 \includegraphics[width = \textwidth]{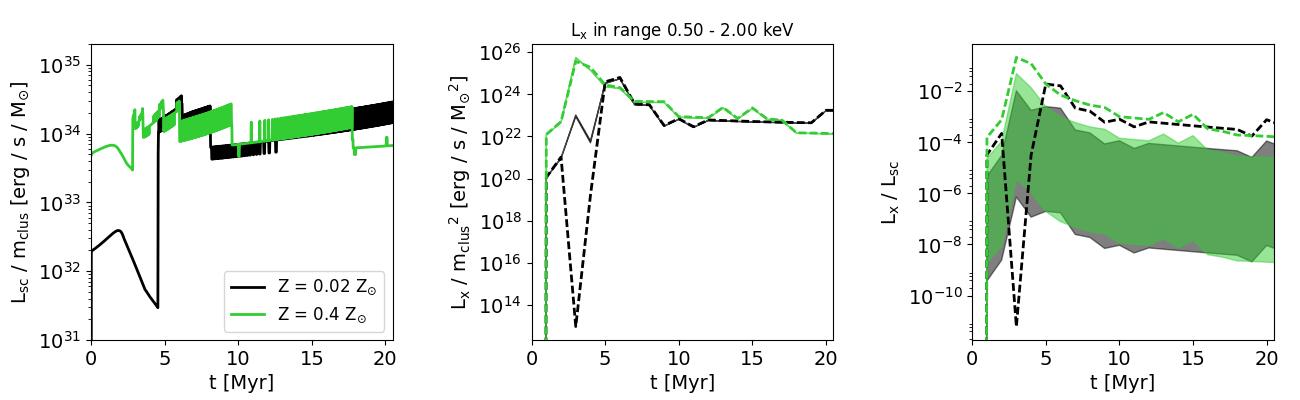}
	\caption{Mechanical energy insertion rate \Lsc\ (\textit{left}), X-ray luminosity \Lx\ in the energy band of \mbox{0.5 -- 2.0~keV} (\textit{middle}), and fraction of the inserted energy transformed into X-ray luminosity from the X-ray-emitting hot gas (\textit{right}). 
	In line with the scaling used in previous works, we normalize \Lsc\ by the cluster mass \mclus, and \Lx\ by $m_{\rm{clus}}^2$ in the left and middle panels, respectively (see Sections~\ref{sec.Xray_Starcluster.LscLx} and \ref{sec.Variation.upperM}).
	The scaling $L_{\rm{X}} \propto m_{\rm{clus}}^2$ is only an approximation, but holds for \mclus\ between $10^3$ and $1.23 \times 10^7$~$M_{\odot}$ (solid lines).  
	The dashed lines in the middle and right panels show the relations for a higher cluster mass of $m_{\rm{clus}} = 8.11 \times 10^7$~$M_{\odot}$, with the lower metallicity (black dashed line) showing a disagreement t\,$\leq$\,4.5\,yr before the onset of supernovae -- see Section~\ref{sec.Xray_Starcluster.LscLx} for details. 
	The shaded regions in the right panel show the evolution of \Lx\,/\,\Lsc\ with time for clusters with masses between 10$^{3}$ (bottom edge) and 1.23\,$\times$\,10$^{7}$~$M_{\odot}$ (top edge).
	Note the zig-zag patterns in the left panel are an artefact caused by interpolation at the end of stellar tracks when the energy from supernovae drops abruptly to zero. In all panels, metallicities of Z\,=\,0.02\,Z$_{\odot}$ and Z\,=\,0.4\,Z$_{\odot}$ are represented by black and green colors, respectively.
	}
	\label{fig.EnIn}
\end{figure*}

Due to stellar winds and supernovae, mechanical energy is transferred to the gas. This can be measured with the energy insertion rate \Lsc\ (also referred as the mechanical luminosity of the cluster), and can be compared to the X-ray luminosity \Lx. 

Figure~\ref{fig.EnIn} shows the evolution of \Lsc\ and \Lx\ as a function of the cluster age (left and middle panel, respectively). The gas density $\rho$ scales linearly with the energy insertion rate \Lsc\ (left panel), and since $\rho \propto m_{\rm{clus}}$ (see top right panel of Figure~\ref{fig.Profiles}), we have $m_{\rm{clus}} \propto L_{\rm{SC}}$ \citep[][]{Silich.etal.2004, Silich.etal.2011, Palous.etal.2013}. Before the onset of supernovae ($t_{\rm{clus}} \leq 2$~Myr for $Z = 0.4$~$Z_{\odot}$ and $t_{\rm{clus}} \leq 4.5$~Myr for $Z = 0.02$~$Z_{\odot}$), \Lsc\ is determined by stellar winds, which depend on the metallicity. Stars with higher metallicities have stronger stellar winds, and thus, their energy insertion rate \Lsc\ is larger. Once supernovae have set in, the energy and mass release are dominated by supernovae, and the metallicity of the gas is only of minor importance to \Lsc. 

The X-ray luminosity \Lx\ (Figure~\ref{fig.EnIn} middle panel) scales with $\rho^2$, as $L_{\rm{X}} \propto n_{\rm{e}} n_{\rm{H}} \Lambda \propto \rho^2$ \citep[see][Equation~3 therein]{Silich.etal.2005, Tenorio-Tagle.etal.2015}. Since $\rho \propto m_{\rm{clus}}$, we find approximately $L_{\rm{X}} \propto m^2_{\rm{clus}}$. This scaling holds for \mclus\ between $10^3$ and $1.23 \times 10^7$~$M_{\odot}$ in our calculations. For larger cluster masses ($m_{\rm{clus}} = 8.11 \times 10^7$~$M_{\odot}$, dashed lines), it breaks down for low metallicity environments at cluster ages between 2 and 4.5~Myr, since the gas inside the clusters become thermally unstable and part of it cools to low temperatures and forms dense warm/cold clumps that do not radiate in X-rays. With the onset of supernovae, all clusters in the considered mass range stabilize, and \Lx\ becomes comparable for both metallicities. The maximum of \Lx\ is reached shortly after the onset of supernovae. This behaviour is consistent with previous results reported in the literature \citep[e.g. ][]{Cervino.etal.2002, Oskinova2005}.

The right panel of Figure~\ref{fig.EnIn} shows which fraction of the inserted energy \Lsc\ is transformed into X-ray luminosity \Lx. The largest efficiency of transforming mechanical energy into X-rays is achieved at $t \sim 3$~Myr for $Z = 0.4$~$Z_{\odot}$, at the same time when \Lx\ has its maximum. For a star cluster at this time with a mass of $m_{\rm{clus}} = 10^7\,M_{\odot}$, we obtain \mbox{\Lx\ / \Lsc\ $\sim 3 \times 10^{-2}$}. Thus, 3\% of the mechanical energy is transformed into soft X-ray emission for this cluster mass. For the low metallicity environment, the peak values of \Lx\ are smaller. A cluster with $m_{\rm{clus}} = 10^7\,M_{\odot}$ at $Z = 0.02$~$Z_{\odot}$ would transform a maximum 0.7\% of its mechanical energy into X-rays.

\subsubsection{X-ray emission of star clusters for a grid of masses and ages}
\label{sec.Xray_Starcluster.grid}
\begin{figure*}
	\includegraphics[width=\textwidth]{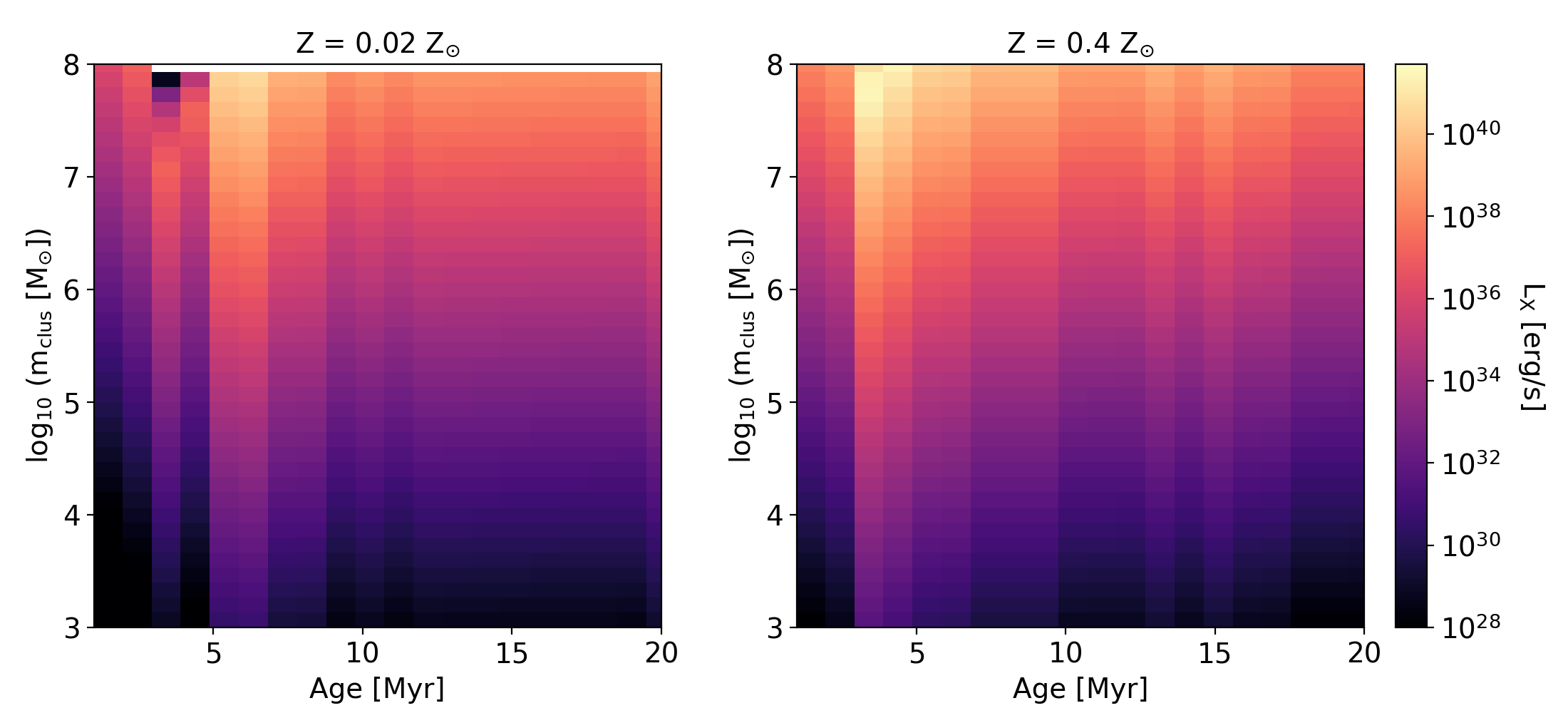}
	\caption{X-ray luminosity \Lx\ of the soft X-ray band (0.5 -- 2.0~keV; color coded) for star clusters in a grid of different masses and ages with metallicities 0.02~$Z_{\odot}$ (\textit{left}) and 0.4~$Z_{\odot}$ (\textit{right}). We use the individual X-ray spectra of clusters to construct the X-ray spectrum of a starburst galaxy (see Section~\ref{sec.galaxies}). In general, the hot gas X-ray emission is dominated by the younger, more massive star clusters.}
	\label{fig.Lx_Lib}
\end{figure*}

We calculate the X-ray spectra for a grid of cluster masses ($10^3$ -- $10^8$~$M_{\odot}$, $\Delta \log_{10}(m_{\rm{clus}} / M_{\odot}) = 0.133$) and ages (0 -- 20~Myr, $\Delta t_{\rm{clus}} = 1$~Myr). This is the basis for constructing the X-ray spectra of starburst galaxies. Figure~\ref{fig.Lx_Lib} shows \Lx\ for the spectra of all clusters for both metallicities (left and right panels, respectively). Young and massive clusters ($t_{\rm{clus}} \sim 3$ -- 6~Myr, $m_{\rm{clus}} \geq 10^7\,M_{\odot}$) show the largest X-ray luminosities \Lx\ up to $10^{41}$~\ergs. In line with the results presented in Figure~\ref{fig.EnIn}, \Lx\ becomes comparable at later times ($t_{\rm{clus}} \gtrsim 4.5$~Myr) for both metallicities -- i.e. once the supernova explosions have set in. Before this, the X-ray emission of star clusters with higher metallicity is larger, since the stellar winds of the high-metallicity stars contain more mass.

We use this grid of X-ray spectra to estimate the X-ray spectrum of a whole starburst galaxy, as described in the following section.

\subsection{Simulation of starburst galaxies composed of star clusters}
\label{sec.galaxies}
Next, we expand our study to starburst galaxies by assuming a given starburst galaxy is composed of star clusters, and estimate the mass \& age distributions of star clusters within such a complex. The mass distribution of star clusters in galaxies can be described by a \textit{cluster mass function}, and was recently studied in observations of the LEGUS dwarf galaxy sample by \citet{Cook.etal.2019}. The authors found a mass function of ${\rm{dN}} / {\rm{d}} m_{\rm{clus}} \propto m_{\rm{clus}}^{\beta}$, with $\beta = -2$\footnote{Note that the index of the cluster mass function $\beta$ is not the same as the index of the initial mass function for stars in the clusters, denoted $\alpha$ in this work.}, in agreement with earlier studies by e.g., \citet{Zhang.Fall.1999, deGrijs.etal.2003, Gieles.Bastian.2008}.

We simulate a starburst galaxy cluster population with a fixed SFR as a function of time up to \mbox{$t_{\rm{gal, max}} = 300$~Myr}. For each time step, we draw star clusters according to the cluster mass function. With proceeding time, the star clusters increase in age, until \mbox{$t_{\rm{clus, max}} = 20$~Myr}, at which point the clusters are discarded. The minimum time step within the galaxy's evolution is set to d$t_{\rm{gal, min}} = 10^3$~yr. In combination with the SFR, this determines the minimum mass of clusters that are drawn in each time step, as
\begin{equation}
\label{eq.mmin}
m_{\rm{min}} = {\rm{d}}t_{\rm{gal, min}} \times \rm{SFR}.    
\end{equation}
Thus, we can draw several clusters within one time step, until the sum of their masses $\sum_i m_{{\rm{clus}}, i} \geq m_{\rm{min}}$. To fulfill the SFR, we draw new clusters according to the mass drawn in the last time step $\sum_i m_{{\rm{clus}}, i}$ and pause it for subsequent time steps if necessary. In this case, the clusters currently present in the starburst galaxy only increase in age. 

Note, that the time evolution depicts the age of the starburst galaxy for fixed SFR, but does not correspond to a real evolutionary time step of the galaxy. Instead, each time step of the galaxy represents a possible realisation of the starburst galaxy, and is therefore of statistical relevance. 

\begin{figure*}
	\includegraphics[width = \textwidth]{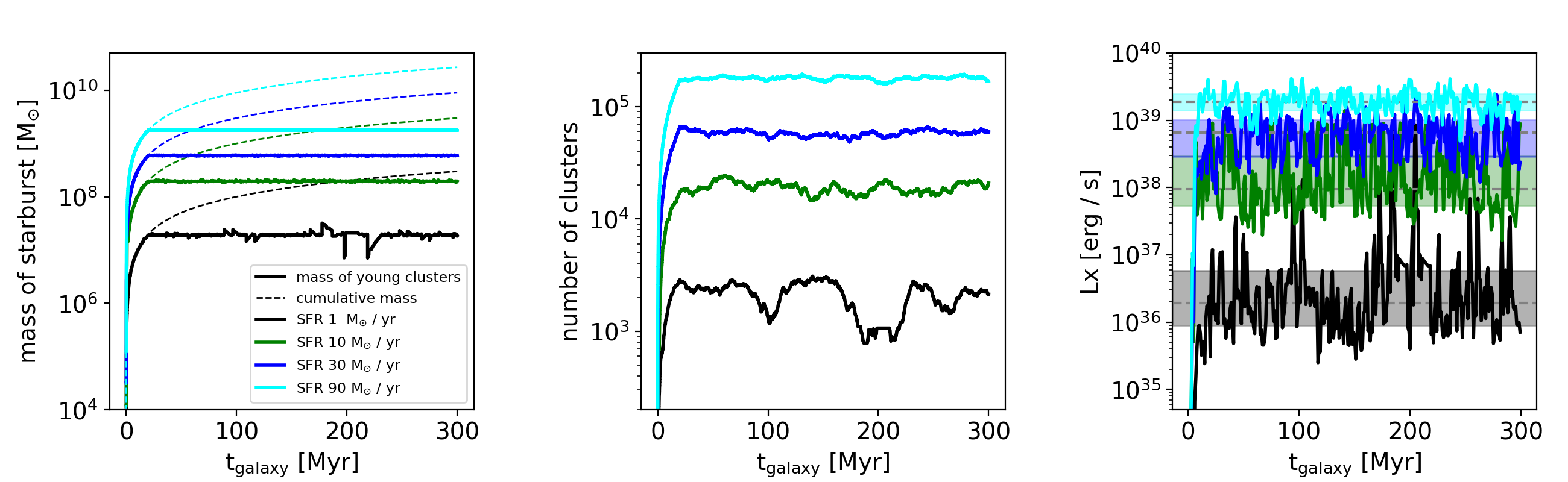}
	\caption{Time evolution of (\textit{left}) the starburst mass (solid lines) and their cumulative masses (dashed lines), (\textit{middle}) the number of clusters for starburst galaxies with different SFRs. We assume here star clusters within a mass range of \mbox{$10^3$ -- $10^7\,M_{\odot}$} and a power law index of the cluster mass function of $\beta = -2$. (\textit{Right}) the X-ray luminosity \Lx\ for $Z = 0.02$~$Z_{\odot}$. The dashed lines and the shaded areas indicate the median values and the range between the 25$^{\rm{th}}$ and 75$^{\rm{th}}$ percentile for each starburst galaxy (see Sections~\ref{sec.galaxies} an \ref{sec.starburst_xray} for details). }
	\label{fig.GalaxyTime}
\end{figure*}
Figure~\ref{fig.GalaxyTime} shows in the left and middle panels how the mass of the starburst galaxy and the number of included clusters changes with time, assuming a constant SFR (color coded). We draw clusters from a cluster mass function with a power law index of $\beta = -2$, while the cluster masses range from $10^3$ -- $10^7\,M_{\odot}$. In the first 20~Myr the full composition of the starburst galaxy is still building up, and its mass increases. After this initial time, we define the starburst galaxy mass as the final mass accumulated within 20~Myr, as set by 
\begin{equation}
\label{eq.galmax}
    m_{\rm{gal, max}} = t_{\rm{clus, max}} \times \rm{SFR}.
\end{equation}
Note, that this mass only contains the contribution of young star clusters, with ages below 20~Myr. The total cumulative stellar mass formed by the starburst is shown by dashed lines in the left panel of Figure~\ref{fig.GalaxyTime}. Comparing it to stellar masses of galaxies in the \citet{Mineo.etal.2012.II} sample ($3\times 10^8$ -- $6\times 10^{10}$\,\MSun), we see that for instance the lowest considered SFR = 0.1\,\Msunyr would have to last $3$\,Gyr to explain the whole stellar mass of the lowest mass galaxy from the sample. On the other hand, the highest considered SFR = 90\,\Msunyr needs 3\,Myr and 670\,Myr to explain the whole stellar mass of the lowest and the highest mass galaxy, respectively.

We further see in Figure~\ref{fig.GalaxyTime} fluctuations in the mass of the starburst galaxy and the number of clusters, for example for the galaxy with $\rm{SFR} = 1$~$M_{\odot}\,\rm{yr}^{-1}$. These fluctuations appear once a massive star cluster was drawn that exceeds the minimum mass $m_{\rm{min}}$ (see Equation~\eqref{eq.mmin}). Thus, to fulfill the SFR, no further star clusters are added to the starburst galaxy in the following time steps, while the age of the clusters continues to increase. The starburst mass as well as the number of star clusters then decreases, causing the fluctuations. As can be seen in the left and middle panel of Figure~\ref{fig.GalaxyTime}, the dips occur after the peaks in the mass of the starburst galaxy.

\subsection{X-ray spectrum of starburst galaxies}
\label{sec.starburst_xray}

\begin{figure*}
	\includegraphics[width = \textwidth]{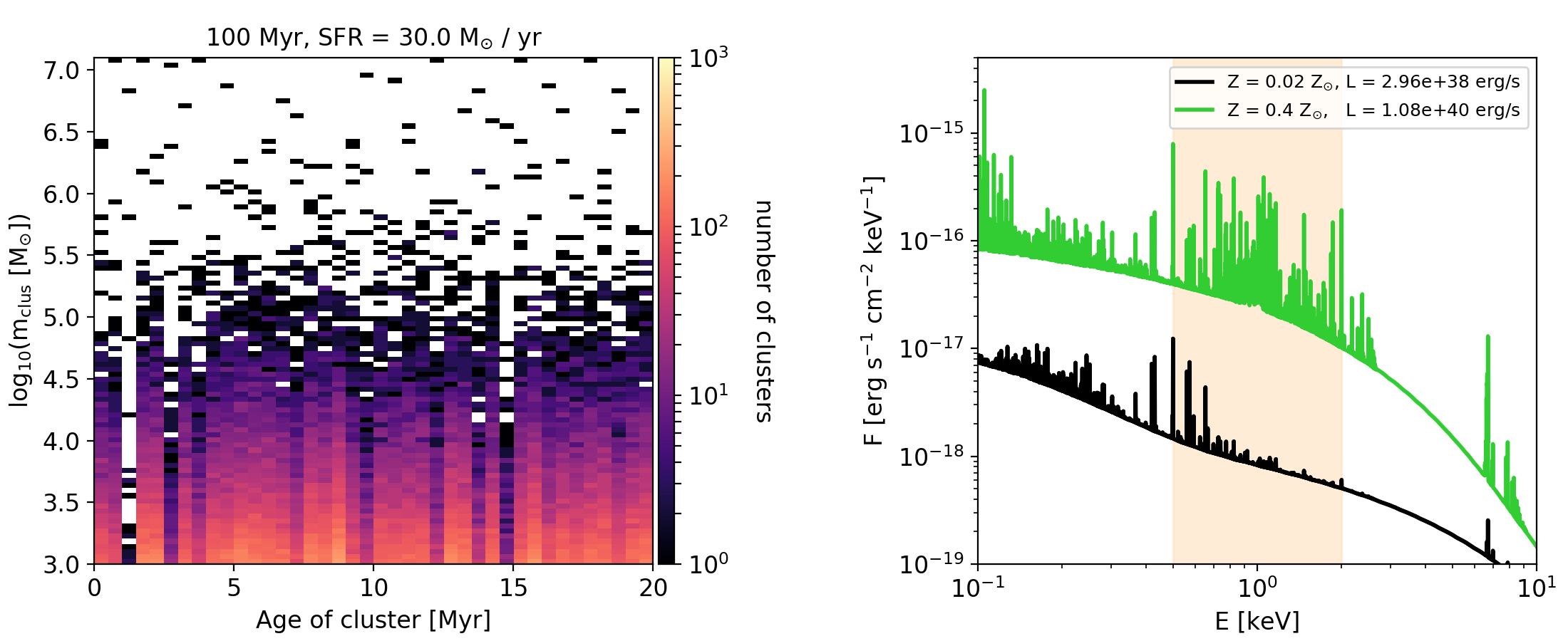}
	\caption{Distribution of star cluster mass and age in one starburst galaxy realization with \mbox{${\rm{SFR}} = 30$~\Msunyr} (\textit{left}). The color bar indicates the number of present clusters. The \textit{right panel} shows the corresponding X-ray spectra for starburst galaxies with each metallicity considered ($Z = 0.02\,Z_{\odot}$ and $0.4\,Z_{\odot}$ in black and green, respectively). The X-ray luminosity in the (shaded) soft X-ray band \mbox{0.5 -- 2.0~keV} for each spectrum is indicated in the legend. }
	\label{fig.GalaxySingle}
\end{figure*}

We calculate the X-ray spectrum of one starburst galaxy by summing the X-ray spectra of the individual star clusters present. Figure~\ref{fig.GalaxySingle} shows the mass and age distribution of star clusters for one starburst galaxy realization with \mbox{$\rm{SFR} = 30$~\Msunyr\ }at a time step of \mbox{$t_{\rm{gal}} = 100$~Myr} (see Figure~\ref{fig.GalaxyTime} for comparison). According to the cluster mass function, most star clusters have masses \mbox{$m_{\rm{clus}} \lesssim 10^5$~$M_{\odot}$}, and only a few are of higher masses. As we have not applied further constraints with respect to the age distribution of the clusters, the mass distribution is fulfilled for all cluster ages. Further, there are age bins with less clusters present. These appear subsequently after we have drawn a high mass cluster in the previous time step (see Section~\ref{sec.galaxies}). 

In the right panel of Figure~\ref{fig.GalaxySingle} we present the X-ray spectra of the corresponding starburst galaxies for both metallicities ($0.02\,Z_{\odot}$ and $0.4\,Z_{\odot}$ in black and green, respectively). In accordance with the spectra of the individual clusters, we find more pronounced emission lines for higher metallicity (see Figure~\ref{fig.Profiles}). The luminosities of the soft X-ray band (shaded area in Figure~\ref{fig.GalaxySingle}) for these two cases are $L_{\rm{X}} \sim 3 \times 10^{38}$~\ergs\ for $0.02\,Z_{\odot}$ and $L_{\rm{X}} \sim 1 \times 10^{40}$~\ergs\ for $0.4\,Z_{\odot}$. Thus, \Lx\ of the starburst galaxy is of a similar order of magnitude as \Lx\ for individual young, massive star clusters (see Figure~\ref{fig.Lx_Lib}). We therefore conclude that the X-ray emission of the hot gas from the whole starburst galaxy can be dominated by the emission from the young, massive star clusters. 

We calculate the X-ray spectra of the starburst galaxies for each time step until $t_{\rm{gal, max}} = 300$~Myr. For the examples of starburst galaxies presented in Figure~\ref{fig.GalaxyTime}, we show the corresponding values of \Lx\ for a metallicity of $0.02\,Z_{\odot}$. As a result of the fluctuations in number of clusters and the changing X-ray emission of the clusters at different ages, \Lx\ of the starburst galaxy likewise fluctuates with time. To characterize \Lx\ of one starburst galaxy statistically, we estimate the median value of \Lx\ in a range of $t_{\rm{gal}}$ between 20 and 300~Myr. We further use the 25$^{\rm{th}}$ and 75$^{\rm{th}}$ percentiles of \Lx\ to determine the range of fluctuation. These values are shown as shaded horizontal bands in Figure~\ref{fig.GalaxyTime}. In the following, we use these measurements for describing \Lx\ in our study.

\section{X-ray emission from hot gas with various assumptions}
\label{sec.results}

\begin{table*}
\caption{Fit of the X-ray luminosity \Lx\ as a function of the SFR for all models. As a fit function we use $L_X = C \times {\rm{SFR}}^{\zeta}$. We fit the data of our models separately for a lower (SFR\,$< 30$~\Msunyr) and upper (SFR\,$\geq 30$~\Msunyr) regime, giving ($C_{1}$, $\zeta_1$) and ($C_2$, $\zeta_2$), respectively. We set the limits between the regimes for each model with the break point in the SFR. }
\label{tab.fit_values}
\begin{tabular}{cccccccc}
\hline
\hline
Met. & $\beta$ & Mass range & Break. point & $C_1$ & $\zeta_1$ & $C_2$ & $\zeta_2$ \\
$Z_{\odot}$ &   & $M_{\odot}$ & $M_{\odot}$ yr$^{-1}$ & \Cunit  &   & \Cunit  &   \\
\hline
0.02 & $-2.0$ & 10$^3$ - 10$^6$ & 3 & $1.50 \times 10^{36}$ & $1.30$ & $(2.20 \pm 0.03) \times 10^{36}$ & $1.01 \pm 0.00$ \\ 
0.02 & $-2.0$ & 10$^3$ - 10$^7$ & 30 & $(2.00 \pm 0.20) \times 10^{36}$ & $1.70 \pm 0.03$ & $(2.20 \pm 0.19) \times 10^{37}$ & $0.99 \pm 0.02$ \\ 
0.02 & $-2.0$ & 10$^3$ - 10$^8$ & 150 & $(4.50 \pm 1.40) \times 10^{36}$ & $1.44 \pm 0.08$ & $2.50 \times 10^{35}$ & $1.96$ \\ 
0.02 & $-2.0$ & 10$^4$ - 10$^7$ & 30 & $(4.50 \pm 0.16) \times 10^{36}$ & $1.54 \pm 0.01$ & $(1.90 \pm 0.11) \times 10^{37}$ & $1.07 \pm 0.01$ \\ 
0.02 & $-2.0$ & 10$^5$ - 10$^7$ & 30 & $(7.10 \pm 0.37) \times 10^{36}$ & $1.52 \pm 0.02$ & $(3.50 \pm 0.26) \times 10^{37}$ & $1.02 \pm 0.01$ \\ 
0.02 & $-1.5$ & 10$^3$ - 10$^7$ & 10 & $(8.10 \pm 0.37) \times 10^{36}$ & $1.88 \pm 0.02$ & $(6.50 \pm 0.22) \times 10^{37}$ & $1.00 \pm 0.01$ \\ 
0.02 & $-1.2$ & 10$^3$ - 10$^7$ & 10 & $(9.00 \pm 0.33) \times 10^{36}$ & $1.93 \pm 0.02$ & $(8.40 \pm 0.26) \times 10^{37}$ & $0.99 \pm 0.01$ \\ 
\hline
0.4 & $-2.0$ & 10$^3$ - 10$^6$ & 3 & $1.00 \times 10^{37}$ & $1.27$ & $(1.70 \pm 0.04) \times 10^{37}$ & $1.00 \pm 0.00$ \\ 
0.4 & $-2.0$ & 10$^3$ - 10$^7$ & 30 & $(1.20 \pm 0.11) \times 10^{37}$ & $1.69 \pm 0.04$ & $(1.40 \pm 0.18) \times 10^{38}$ & $1.02 \pm 0.02$ \\ 
0.4 & $-2.0$ & 10$^3$ - 10$^8$ & 150 & $(2.30 \pm 0.47) \times 10^{37}$ & $1.67 \pm 0.05$ & $8.40 \times 10^{36}$ & $1.80$ \\ 
0.4 & $-2.0$ & 10$^4$ - 10$^7$ & 30 & $(2.30 \pm 0.30) \times 10^{37}$ & $1.60 \pm 0.05$ & $(1.70 \pm 0.14) \times 10^{38}$ & $1.04 \pm 0.02$ \\ 
0.4 & $-2.0$ & 10$^5$ - 10$^7$ & 30 & $(3.90 \pm 0.56) \times 10^{37}$ & $1.60 \pm 0.05$ & $(2.60 \pm 0.22) \times 10^{38}$ & $1.02 \pm 0.02$ \\ 
0.4 & $-1.5$ & 10$^3$ - 10$^7$ & 10 & $(4.80 \pm 0.56) \times 10^{37}$ & $1.80 \pm 0.07$ & $(4.80 \pm 0.07) \times 10^{38}$ & $1.00 \pm 0.00$ \\ 
0.4 & $-1.2$ & 10$^3$ - 10$^7$ & 10 & $(5.20 \pm 0.21) \times 10^{37}$ & $1.99 \pm 0.02$ & $(6.40 \pm 0.28) \times 10^{38}$ & $0.99 \pm 0.01$ \\ 
\hline
\end{tabular}
\end{table*}

\subsection{\Lx\ as a function of the SFR}
\label{sec.results.varSFR}
\begin{figure}
    \includegraphics[width = 80mm]{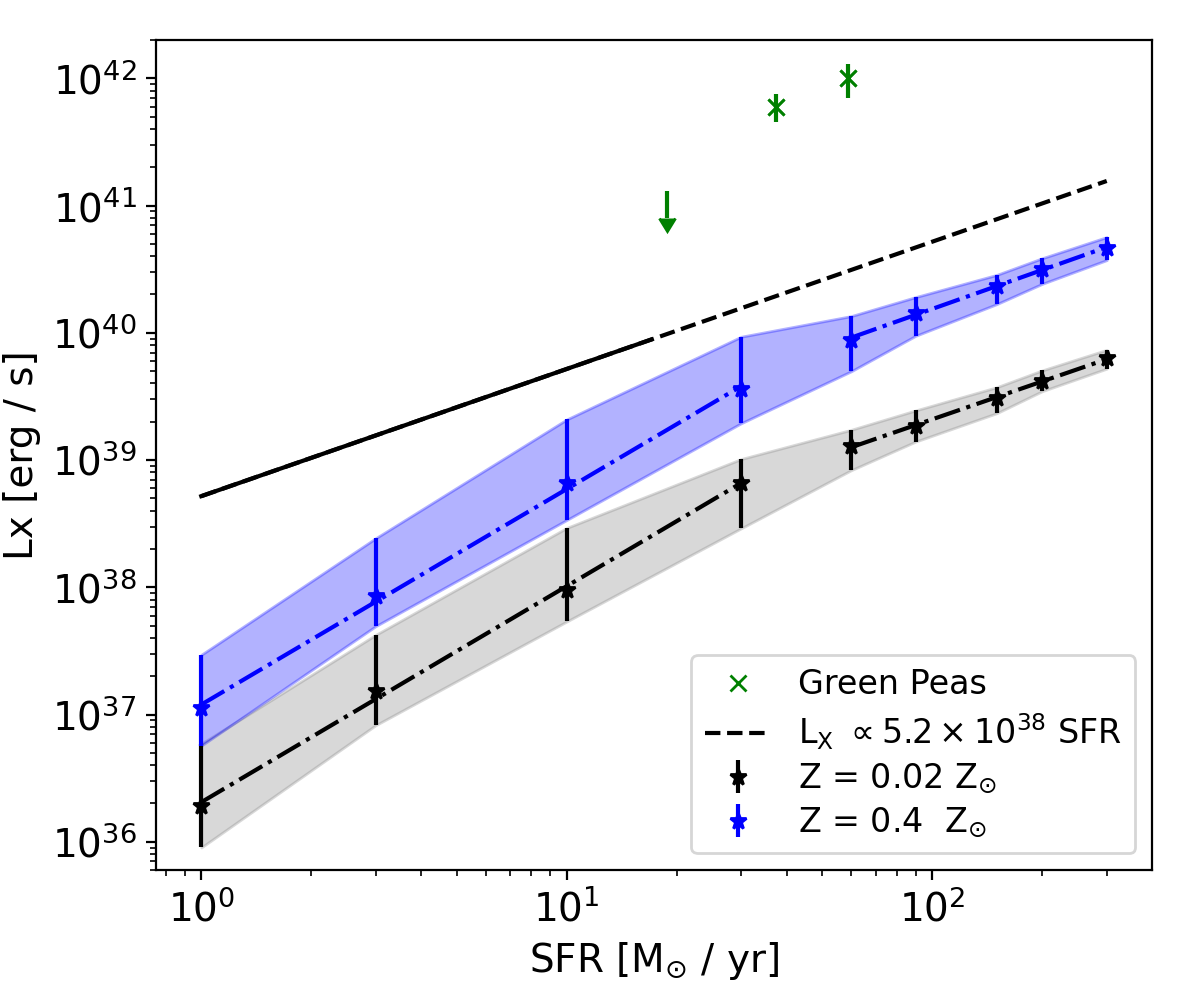}
    \caption{X-ray luminosity (\Lx) of the starburst galaxies as a function of their SFR, assuming different metallicities (color coded). The starburst galaxies consist of star clusters with masses in the range of $10^3$ -- $10^7$~$M_{\odot}$, distributed according to a cluster mass function with power law index $\beta = -2$. In general, \Lx\ is larger for higher metallicities. The dashed-dotted lines are the fits to our data (see text). For comparison, the dashed line shows the observationally obtained relation between \Lx\ and the SFR for star-forming galaxies from \citet{Mineo.etal.2012.II}, and the green symbols indicate \Lx\ for the Green Pea galaxies from \citet{Svoboda.etal.2019}.}
    \label{fig.Result_SFR_met}
\end{figure}
As a first approach, we analyse the soft X-ray emission of starburst galaxies composed of star clusters with masses between $10^3$ to $10^7$~$M_{\odot}$, and a cluster mass function with power law index $\beta = -2$. Figure~\ref{fig.Result_SFR_met} shows \Lx\ of these starburst galaxies as a function of the SFR for different metallicities. The points indicate the median values of \Lx, and the shaded areas the range between the 25$^{\rm{th}}$ and 75$^{\rm{th}}$ percentiles (see Section~\ref{sec.starburst_xray}). In general, \Lx\ increases with increasing SFR and in line with the previous results, we find \Lx\ to be larger for starburst galaxies at higher metallicities. For comparison, we plot the observed relation between hot gas X-ray luminosity and SFR from \citet{Mineo.etal.2012.II}; $L_{0.5 - 2 {\rm keV}} / \textrm{SFR} \sim (5.2 \pm 0.2) \times 10^{38}$~\Cunit\ which was obtained by fitting X-ray spectra with the \mekal\ model inside \xspec. In contrast to the results of \citeauthor{Mineo.etal.2012.II}, our results cannot be described by one single power law. Instead, our relation between \Lx\ and the SFR consists of two ranges, in which the range of smaller SFRs ($< 30$~\Msunyr) has a steeper index, and the range of larger SFRs ($\geq 30$~\Msunyr) a flatter one. We therefore fit our data separately for these two ranges, each by a function of
\begin{equation}
    \label{eq.fit}
    L_X = C \times {\rm{SFR}}^{\zeta}
\end{equation}
The results of all fits are summarized in Table~\ref{tab.fit_values}. We indicate by the breaking point the limit between the two SFR ranges, which is 30~\Msunyr in this case. We find $\zeta_1 \sim 1.7$ for the range of lower SFRs for both metallicities, and $\zeta_2 \sim 1$ for the range of larger SFRs. Thus, our normalization $C$ can only be compared to the results of \citet{Mineo.etal.2012.II} for the range of larger SFRs. Here our data results in $C_2 \sim 2.20 \times 10^{37}$ and $\sim 1.40 \times 10^{38}$~\Cunit\ for the metallicities of $Z = 0.02\,Z_{\odot}$ and $0.4\,Z_{\odot}$, respectively. These indices are about a factor of $24$ and $4$ times smaller than the index found by \citet{Mineo.etal.2012.II}. We additionally note that for SFRs above 17~\Msunyr, the relation of \citeauthor{Mineo.etal.2012.II} that we compare to is an extrapolation. \\

In Figure~\ref{fig.Result_SFR_met}, the green points indicate \Lx\ as measured for the three Green Pea galaxies in \citet{Svoboda.etal.2019}. As the observed X-ray emission is even higher for the Green Pea galaxies than the relation by \citet{Mineo.etal.2012.II}, our model predicts even less of the X-ray emission to originate from the hot gas. However, our results are obtained on a basis of several assumptions/restrictions, such as the star cluster's mass range and the power law index of the cluster mass function $\beta$, as well as the fixed core and cutoff radii of the star clusters. In addition, the inclusion of chemically homogeneous evolving stars could have an impact on our results. Thus in the following sections, we test how these factors influence the value and scaling of \Lx.

\subsection{Dependence on star cluster masses}
\label{sec.Variation}
\begin{figure*}
    \includegraphics[width = \textwidth]{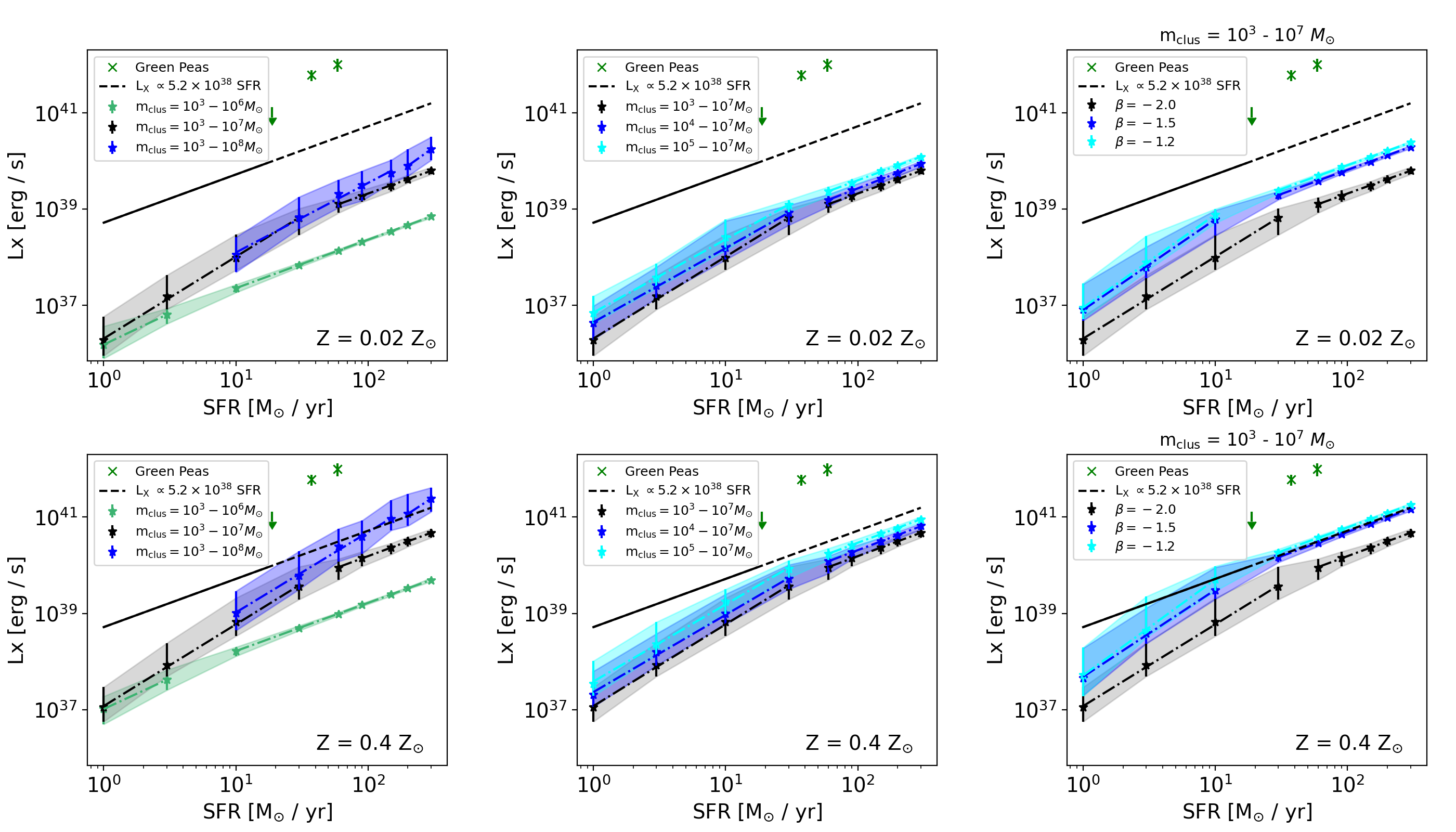}
    \caption{Similar to Figure~\ref{fig.Result_SFR_met}, we show how different assumptions in constructing the starburst galaxies influence the relation between \Lx\ and SFR. We vary the upper \textit{(left)} and lower mass limit \textit{(middle)} of the cluster's mass range with the power law index of the cluster mass function $\beta = -2$. In contrast we vary $\beta$ in the \textit{(right)} panels whilst $m_{\rm{clus}} \sim 10^3$ -- $10^7$~$M_{\odot}$. The rows represent the metallicities of $Z = 0.02\,Z_{\odot}$ \textit{(upper)} and $Z = 0.4\,Z_{\odot}$ \textit{(lower)}. \Lx\ increases if the number of massive star clusters increases and/or star clusters with larger masses are present. This is the case for a mass range shifted towards higher cluster masses, and for flatter indices $\beta$ (see Section~\ref{sec.Variation}).}
    \label{fig.Res_Parameters}
\end{figure*}

The star cluster's mass range in galaxies is an ongoing matter of debate. Young massive clusters and super star clusters are observationally found to have masses between $10^5$ and $10^8$~$M_{\odot}$ \citep[e.g.][]{Whitmore.2003, OConnell.2004, Bastian.etal.2006, Reines.etal.2008}. Recently, \citet{Norris.etal.2019} provided a catalogue of compact stellar systems and argue in favour of a fundamental maximum star cluster mass of $\sim 5 \times 10^7\,M_{\odot}$, requiring an initial cluster mass of $10^8\,M_{\odot}$. The low mass end of the star cluster's mass range is hard to constrain due to the lower luminosities (see e.g. Figure~\ref{fig.GalaxySingle}). However, such clusters enclose the mass in the galaxy and thus have an influence on the amount of mass left for more massive star clusters. Various observational studies investigated the distribution of cluster masses; \citet{Adamo.etal.2010.I} find in Haro~11 more than 30\% of the clusters to have masses $> 10^5\,M_{\odot}$, while the fraction of low-mass and very young clusters ($< 10^4\,M_{\odot}$, \mbox{1 -- 3~Myr}) is missing. Likewise, in SBS~0335 only clusters with high masses ($10^5$ to $10^6\,M_{\odot}$) are present \citep{Adamo.etal.2010.II}. Thus next we study the dependence of cluster mass range and power law index on the distribution functions we derive (see Figure~\ref{fig.Res_Parameters}). Whilst the individual clusters in Green Pea galaxies are unresolved, we assume the structure could be similar to Haro~11. Furthermore some of the parameters found by \citeauthor{Adamo.etal.2010.I} appear extreme compared to regular starburst galaxies but should be explored in the case of Green Pea galaxies which also appear extreme.

\subsubsection{Varying the upper limit of the star cluster mass range}
\label{sec.Variation.upperM}
In Figure~\ref{fig.Res_Parameters} we show in the left panels the relation between \Lx\ and the SFR, assuming different upper limits on the cluster mass range, while keeping the lower limit constant at $10^3$~$M_{\odot}$. For comparison with observations, we plot the relation found from \citet{Mineo.etal.2012.II}, and the X-ray luminosity measurements of the Green Pea galaxies from \citet{Svoboda.etal.2019}. If we decrease the upper mass of the cluster's mass range to $10^6\,M_{\odot}$ (light green), the X-ray luminosity \Lx\ becomes smaller compared to the cases with larger upper masses for all SFRs. This is a result of the absence of high mass star clusters in these starburst galaxies. Although the total mass of the starburst galaxy is fixed by the SFR (see Equation~\eqref{eq.galmax}), the distribution of cluster masses changes with the upper limit of the cluster mass range. Therefore, as for a single star cluster we find approximately $L_{\rm{X}} \propto m_{\rm{clus}}^2$ (see Section~\ref{sec.Xray_Starcluster.LscLx}), as the absence of high mass star clusters in the starburst galaxy leads to a decrease of \Lx. Similarly, \Lx\ increases when we set the star cluster mass range to $10^3$ -- $10^8$~$M_{\odot}$.

Further, \Lx\ fluctuations decrease with increasing SFR (denoted by the shaded area around the points). The higher the SFR, the more star clusters at the high mass end of the mass range that can coexist. Since the fluctuations arise from the appearance of massive clusters, once the number of high mass clusters stabilizes, the fluctuations decrease. This happens faster for a lower upper cluster mass of $10^6$~$M_{\odot}$. 

For a cluster mass range of $10^3$ -- $10^8$~$M_{\odot}$ we only calculate \Lx\ for $\rm{SFR} \geq 10$~\Msunyr, as otherwise the mass of the most massive cluster is larger than the mass of the total starburst galaxy $m_{\rm{gal, max}}$ (see Equation~\eqref{eq.galmax}). In this particular case, a starburst galaxy at $\rm{SFR} = 1$~\Msunyr, has $m_{\rm{gal, max}} = 2 \times 10^7\,M_{\odot}$ according to Equation~\eqref{eq.galmax}, and thus, cannot be represented when including cluster masses up to $10^8$~$M_{\odot}$. This is why the points of low SFRs are missing for this cluster mass range in Figure~\ref{fig.Res_Parameters}.

We compare the two metallicities in the upper and lower rows of Figure~\ref{fig.Res_Parameters}. Qualitatively, \Lx\ scales similarly with the SFR for both metallicities. Quantitatively, \Lx\ is systematically below the observational scaling by \citet{Mineo.etal.2012.II}, and the X-ray emission of the Green Pea galaxies for the panels with $Z \sim 0.02$~$Z_{\odot}$. For $Z = 0.4$~$Z_{\odot}$ and a cluster mass range of $10^3$ -- $10^8$~$M_{\odot}$ however \Lx\ is consistent with the observational relation by \citet{Mineo.etal.2012.II} within an order of magnitude, which appears to be driven by increasing the number of higher-mass clusters in a given simulation. 

As above, we fit the relation between \Lx\ and SFR with a broken power law for two ranges of SFR (Equation~\eqref{eq.fit}). The breaking points separating these two regimes differ with the cluster mass ranges, and are 3~\Msunyr\ for a cluster mass range of $10^3$ -- $10^6\,M_{\odot}$, and 150~\Msunyr\ for a cluster mass range of $10^3$ -- $10^8\,M_{\odot}$. 
In the range of lower SFR, the relation between SFR and \Lx\ is determined by the number of clusters, and by the mass of the most massive clusters. 
In the range of larger SFR, the relation depends only on the number of clusters present, as several of the most massive clusters are at the upper mass limit. Thus the breaking point shifts with the upper limit of the cluster mass range. For all plots shown in the left panels, we find the range of low SFR $\zeta_1 > 1$, while in the range of high SFR $\zeta_2 \sim 1$ (see Table~\ref{tab.fit_values}). Thus, only the upper range of SFR is comparable with the relation by \citet{Mineo.etal.2012.II}.

For the mass range of $10^3$ -- $10^6\,M_{\odot}$ we find normalizations $C_2$ of $2.2 \times 10^{36}$ and $1.7 \times 10^{37}$~\Cunit\ for the metallicities of $0.02$ and $0.4\,Z_{\odot}$, respectively -- i.e. a factor of 236 and 30 times smaller than the index reported by \citet{Mineo.etal.2012.II}.

\subsubsection{Varying the lower limit of the star cluster mass range}
\label{sec.Variation.lowerM}
In Figure~\ref{fig.Res_Parameters} we show the relation between \Lx\ and SFR by varying the lower limit of the cluster mass range, and keeping the upper limit at $10^7\,M_{\odot}$ (middle panels). By increasing the lower mass limit from $10^3$ to $10^5\,M_{\odot}$, the X-ray luminosity \Lx\ increases, as the available mass in the starburst galaxy is distributed to clusters of higher mass. The increase of \Lx\ is smaller compared to the case where we varied the upper mass limit, as in the individual star clusters the X-ray emission scales with $m_{\rm{clus}}^2$ (see Sections~\ref{sec.Xray_Starcluster.LscLx} and \ref{sec.Variation.upperM}). Although for the higher metallicity we find larger values for \Lx\ (bottom middle panel), we do not reach the values expected from observations by \citet{Mineo.etal.2012.II} nor by \citet{Svoboda.etal.2019}. For the fits with Equation~\eqref{eq.fit} the breaking points between the two ranges in SFR do not change in this case, because the upper mass of the mass range is constant, and the mass distributed on lower mass scales remains similar. For the upper SFR range, $\zeta_2 \sim 1$ for all cases, and thus, comparable with the relation found by \citet{Mineo.etal.2012.II}. The indices of the fits in the upper SFR regimes are between $\sim$2.0 and $\sim$3.5$\times 10^{37}$~\Cunit\ for a metallicity of 0.02~$Z_{\odot}$, and between $\sim$1.4 and $\sim$2.6$\times 10^{38}$~\Cunit\ for a metallicity of 0.4~$Z_{\odot}$ (see Table~\ref{tab.fit_values}). Comparing with \citet{Mineo.etal.2012.II}, these values correspond to $\leq 1/15$ and $\leq 1/2$ of the index found in their work for the two metallicities.

\subsection{Varying the power law index of the cluster mass function}
\label{sec.Variation.beta}

As introduced in Section~\ref{sec.galaxies}, the distribution of cluster masses within the given range is described by the cluster mass function with a power law index $\beta$. In accordance with e.g. \citet{Zhang.Fall.1999, deGrijs.etal.2003, Gieles.Bastian.2008} we set $\beta = -2$ in the analyses described above. However, it is debated whether $\beta$ varies with the type of galaxy or SFR. For example in the LEGUS dwarf galaxy sample with SFRs between 0.005 to $0.5$~\Msunyr\ \citet{Cook.etal.2019} find $\beta = -2$, whereas in NGC~4449 with ${\rm{SFR}} \sim 1.5$~\Msunyr\ \citet{Annibali.etal.2011} report a flatter power law index of $\beta \sim -1.5$. Likewise, \citet{Adamo.etal.2010.I} find the index is found to be flatter in Haro~11 with a current SFR of $22 \pm 3$~\Msunyr, as observationally indicated by the luminosity function from the star clusters \citep{Adamo.etal.2010.I}. With the luminosity function, the cluster mass function is expected to change, as they are directly connected to the star formation processes \citep{Adamo.etal.2010.I}. 

Since we construct starburst galaxies at higher SFRs in our simulations, we analyse how a flatter power law index of the cluster mass function influences the scaling between \Lx\ and SFR. The results are shown in the right panels of Figure~\ref{fig.Res_Parameters}. We assume a cluster mass range of \mbox{$10^3$ -- $10^7\,M_{\odot}$}, and vary the cluster mass function $\beta$ between $-2.0$, $-1.5$, $-1.2$. For flatter indices we find an increase in \Lx\ as well as smaller fluctuations of \Lx\ around the median value. This is a result of the increase in number of high mass clusters. The flatter the power law index, the more high mass clusters are present, which dominate the X-ray luminosity of the whole starburst galaxy. Comparing results for the two metallicities (upper and lower right panels), we find \Lx\ to be in agreement with the observational relation of \citet{Mineo.etal.2012.II} for $Z = 0.4$~$Z_{\odot}$ and $\beta \geq -1.5$. The fit in the upper SFR range for $\beta = -1.2$ has a normalization of $C_2 \sim 8.4 \times 10^{37}$ and $\sim 6.4 \times 10^{38}$~\Cunit\ for the two metallicities, respectively, which corresponds to 1/6 and 1/23 of the index found by \citet{Mineo.etal.2012.II}. However, none of our simulations reproduces the luminosities observed for the Green Pea galaxies as reported in \citet{Svoboda.etal.2019}. 

\subsection{Varying the core and cutoff radii}
\label{sec.Variation.radii}
\begin{figure}
    \centering
    \includegraphics[width = 80mm]{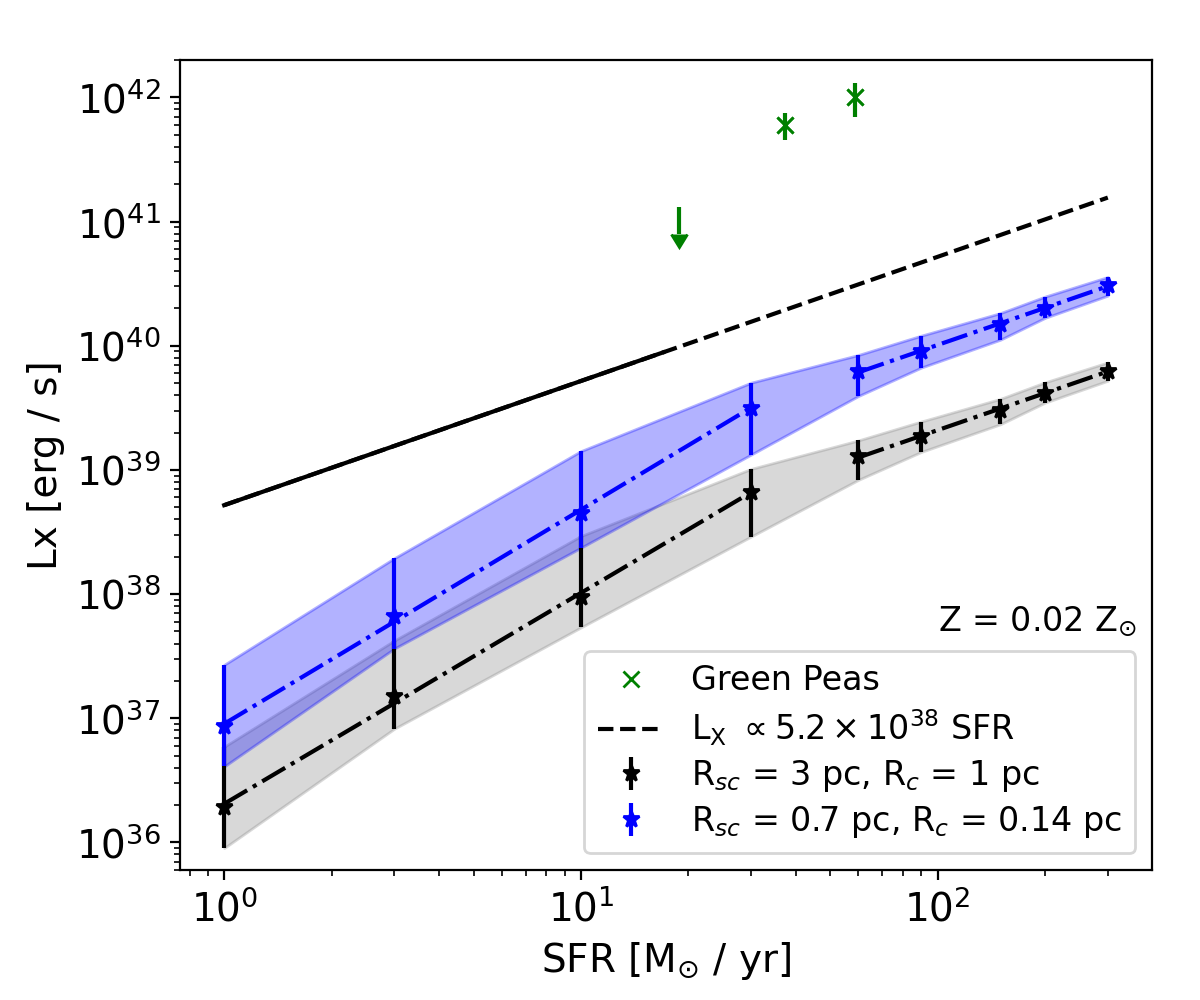}
    \caption{Similar to Figure~\ref{fig.Result_SFR_met}, we show for a metallicity of $0.02$~$Z_{\odot}$ the relation between \Lx\ and SFR for starburst galaxies with smaller cutoff and core radii \Rsc\ and \Rc, as found by \citet{Espinoza.etal.2009} in the Arches cluster (blue symbols). The black symbols represent the same values as shown in Figure~\ref{fig.Result_SFR_met}. Smaller radii lead to an increased X-ray luminosity, but \Lx\ is still beyond the observed values.}
    \label{fig.Result_SFR_RscRc}
\end{figure}

So far we have assumed the core radius \Rc\ and cutoff radius \Rsc\ of the star cluster to be $R_{\rm{C}} = 1$~pc, and $R_{\rm{SC}} = 3$~pc, respectively (see Section~\ref{sec.starcluster.diststars}). However, some star clusters have been found to be more compact than this. \citet{Espinoza.etal.2009} studied the Arches cluster with a mass of $\sim 2 \times 10^4$~$M_{\odot}$, finding \mbox{$R_{\rm{C}} = 0.14$~pc}, and \mbox{$R_{\rm{SC}} = 0.7$~pc} (see their Table~8). With these values, the Arches cluster is one of the densest known young clusters in the Milky Way. The smaller radii lead to higher concentrations of mass in the cluster, which leads to increased gas density, temperature and hence X-ray emission. 

We study the influence of \Rc\ and \Rsc\ on our results by repeating the calculation of \Lx\ as a function of SFR for starburst galaxies with cluster masses \mbox{$10^3$ -- $10^7$~$M_{\odot}$}, a cluster mass function power law index of $\beta = -2$, and \Rc\ and \Rsc\ set to the values found by \citet{Espinoza.etal.2009}. Figure~\ref{fig.Result_SFR_RscRc} shows the resulting relation derived after decreasing the cluster radii, which shifts the X-ray luminosity on average to higher values. However even with these assumptions \Lx\ is still smaller than the observational results from \citet{Mineo.etal.2012.II} and \citet{Svoboda.etal.2019}.

\subsection{Including chemically homogeneous stars}
\label{sec.results_CHE}
\begin{figure}
    \centering
    \includegraphics[width = 80mm]{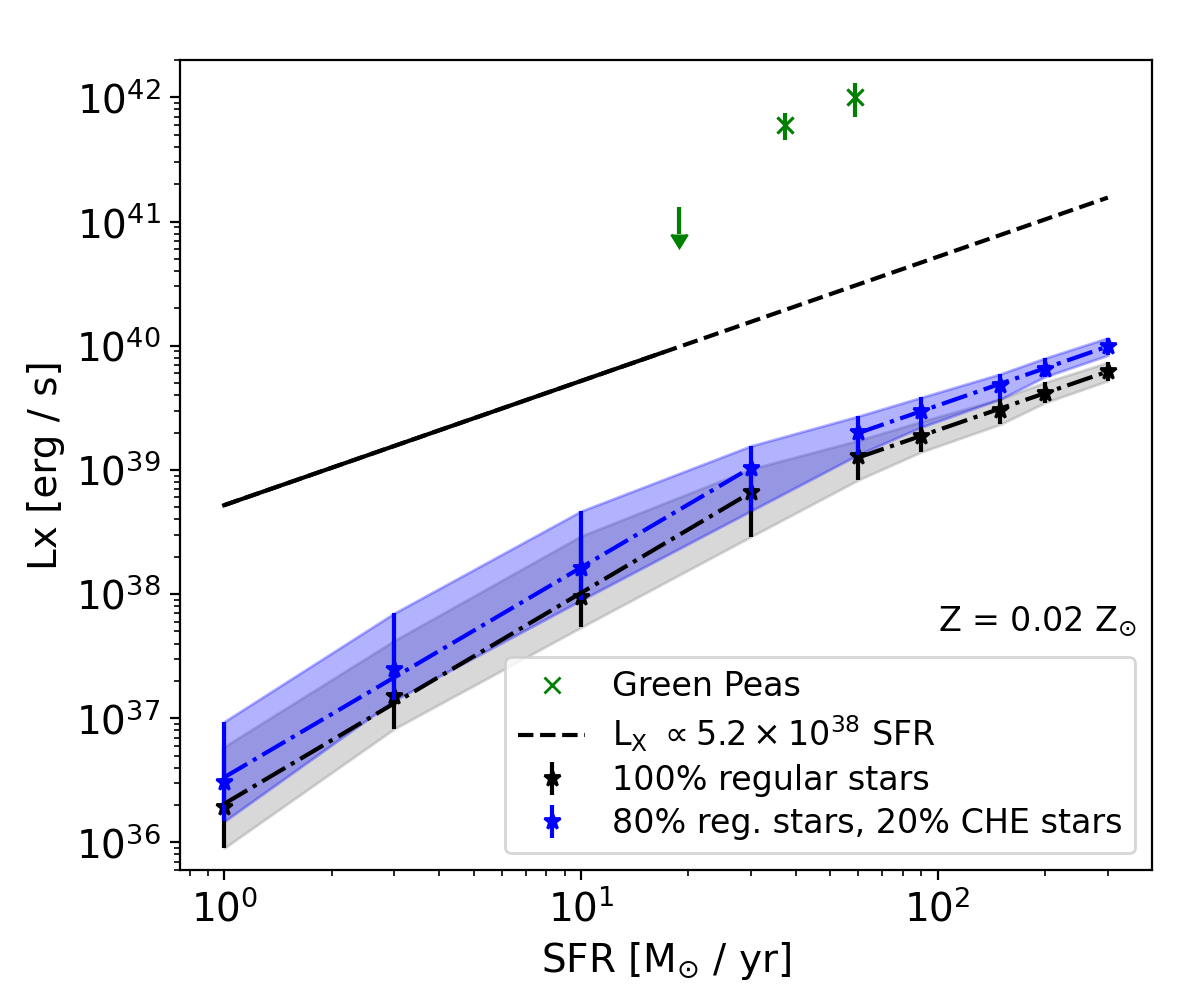}
    \caption{Similar to Figure~\ref{fig.Result_SFR_met}, we show for a metallicity of $0.02$~$Z_{\odot}$ the relation between \Lx\ and the SFR for starburst galaxies assuming that 20\% of the stars are chemically homogeneous evolving stars (CHE, blue symbols). The black symbols represent the same values as shown in Figure~\ref{fig.Result_SFR_met}. By including chemically-homogeneously evolving stars, \Lx\ increases slightly. However, it does not reach the X-ray emission expected from observations.}
    \label{fig.Result_SFR_chem}
\end{figure}

Star clusters with gas at low-metallicity can host fast rotating stars, which may evolve quasi chemically-homogeneously \citep{Yoon.etal.2006, Brott.etal.2011, Szecsi.etal.2015, Szecsi.conf.2015, Szecsi.conf.2017, Garcia.etal.2019}. The temperatures of such stars are higher than of regular stars, and some may have strong winds (e.g., \citealt{Kubatova.etal.2019}). Here we test whether the hot stellar winds of these chemically-homogeneously evolving stars would contribute significantly to the X-ray emission. We apply the hot chemically-homogeneously evolving stellar models from the BoOST project \citep{Szecsi.etal.2020} with low metallicity (Z = 0.02~$Z_{\odot}$), to obtain the mass loss in each star cluster. For these models, a correction for the wind optical depth is taken into account, following the method presented in Chapter~4.5.1 of \citet{Szecsi.2016}. Based on observational comparisons to the star-forming dwarf galaxy IZw18, \citet{Szecsi.etal.2015} find that $\sim$20\% of the stellar population is likely evolving homogeneously. Therefore, for our low-metallicity cluster (Z = 0.02~$Z_{\odot}$) we combine mass loss from 80\% regular and 20\% chemically-homogeneously evolving stars when constructing our star clusters, and calculate their X-ray emission for a grid of masses (\mbox{$10^3$ -- $10^7$~$M_{\odot}$}) and ages ($\leq 20$~Myr) in the same way as in Section~\ref{sec.galaxies}. From Figure~\ref{fig.Result_SFR_chem}, we find the chemically-homogeneously evolving stars to increase \Lx\ slightly, but still significantly below the X-ray luminosities measured for the Green Pea galaxies.

\section{Discussion}
\label{sec.discussion}

\subsection{Comparison with previous work}

The X-ray emission from hot gas in the interstellar medium from starburst galaxies was previously simulated in a population synthesis model by \citet{Cervino.etal.2002}. The authors assumed a single starburst galaxy and derived the energy output from every star using stellar evolution models. Our starburst galaxies are composed of several star clusters. For modelling the stars, \citeauthor{Cervino.etal.2002} assumed a Salpeter initial mass function, and considered stars with masses between 2 and 120~$M_{\odot}$, parabolic interpolation between the Geneva stellar models \citep{Schaller.etal.1992, Meynet.eta.1994} and supernova explosions. The X-ray emission is then a result of the contribution from supernova remnants and the hot gas. However, their approach of calculating the hot gas properties of a starburst galaxy differs significantly from ours. Instead of computing the wind dynamics, \citeauthor{Cervino.etal.2002} define an efficiency parameter $\epsilon_{\rm{eff}}^{\rm{x}}$ describing the variable fraction of total mechanical energy that contributes to the heating of the diffuse interstellar medium. 

\citet{Cervino.etal.2002} and \citet{Oti-Floranes2012, Oti-Floranes2014} use this model to find a good set of parameters to explain observations by first constraining the age of the observed object, then the corresponding kinetic energy released to the gas at this time, and adopt the mass transformed into stars and efficiency parameter $\epsilon_{\rm{eff}}^{\rm{x}}$ required to fit the observation. The authors typically find efficiencies from 0.4\% up to 10\% for star-forming objects \citep[see e.g.][]{Summers.etal.2001, Summers.etal.2004, Mas-Hesse.etal.2008, Oti-Floranes2012}. This is in line with our models; as we have shown in Section~\ref{sec.Xray_Starcluster.LscLx}, $\lesssim 3$\% of the energy insertion rate in a star cluster is transformed into X-rays. The actual value depends on the metallicity as well as the mass and evolutionary state of the star cluster. 

\subsection{Comparison with observations}
The X-ray emission of a sample of nearby star-forming galaxies was studied in a series of papers by \citet{Mineo.etal.2012.I, Mineo.etal.2012.II, Mineo.etal.2014}. The galaxies considered had SFRs of $\sim 0.1$ -- $17$~\Msunyr. To infer the gas emission, \citet{Mineo.etal.2012.II} measured the total X-ray flux of each galaxy and subtracted the flux of resolved and unresolved faint High-Mass X-ray Binaries. In addition, the authors attempt to correct for unresolved sources of X-ray emission from Low-Mass X-ray Binaries, young stellar objects and cataclysmic variables. However, the contributions of both are found to be negligible and the models estimating their contribution to be too uncertain to perform an accurate subtraction. Therefore, the X-ray luminosity of the diffuse interstellar medium (0.5 -- 2.0~keV) presented by \citet{Mineo.etal.2012.II} needs to be considered as an upper limit. 

\citet{Mineo.etal.2012.II} distinguish two components in the unresolved emission: the diffuse X-ray emission and the X-ray emission of the thermal component (modelled with the \mekal\ model in \xspec). Here, the diffuse emission refers to the contribution of sources with hard spectra (e.g., unresolved faint compact sources and/or hotter gas) that could not be reproduced well from the lower-temperature \mekal\ components. The authors find linear relations with indices of $L_{0.5 - 2\,\rm{keV}}^{\rm{diff}} / {\rm SFR} \sim (8.3 \pm 0.1) \times 10^{38}\,$\Cunit\ and $L_{0.5 - 2\,\rm{keV}}^{\rm{mekal}} / {\rm SFR} \sim (5.2 \pm 0.1) \times 10^{38}\,$\Cunit. We compared the relation between the X-ray luminosity and the SFR of our work to the \mekal\ component of \citet{Mineo.etal.2012.II}. In general, the X-ray luminosity of our models varies when assuming different ranges of cluster masses within the starburst galaxy, and assuming different power law indices $\beta$ of the cluster mass function. For the widely reported value of $\beta = -2$ and \mclus\ in the range of \mbox{$10^3$ -- $10^7$~$M_{\odot}$} we find our X-ray emission to be smaller compared to the emission expected from e.g., \citeauthor{Mineo.etal.2012.II}. However, if we allow more massive clusters to be present by taking a larger mass range (\mbox{$m_{\rm{clus}} \sim 10^3$ -- $10^8$~$M_{\odot}$}) or by adopting a flatter power law index of $\beta \geq -1.5$, as suggested by some studies of dwarf galaxies \citep[e.g.][]{Adamo.etal.2010.I, Annibali.etal.2011}, the X-ray emission of our models for $Z = 0.4$~$Z_{\odot}$ (see Figure~\ref{fig.Res_Parameters}) is in line with the expectations by \citet{Mineo.etal.2012.II}. 

\subsection{Limitations of our model}

The model we designed for constructing starburst galaxies assumes stars form in individual star clusters within the starburst galaxy. We demonstrated how the star cluster mass range and cluster mass function power law index $\beta$ influences the X-ray luminosity of the hot gas from the colliding stellar winds. Here we discuss additional aspects influencing the X-ray emission from the hot gas.

When calculating the X-ray emission of the hot gas in the interstellar medium, we have accounted for the collision of nearby stellar winds, but the interaction with the interstellar medium surrounding the star clusters has been neglected. We have also not modeled the presence of clouds exposed to the galactic-scale wind, as done by \citet{Marcolini2005}. Such effects are mainly important for massive clusters and high density environments, and could be taken into account in future studies.

For each star cluster, we assume the same core radius \Rc\ and cutoff radius \Rsc, independent of the cluster mass (see Section~\ref{sec.starcluster}). Therefore, the volume of each cluster is the same, and the gas density in the clusters increases with increasing cluster mass. We have chosen moderate values of $R_{\rm{C}} = 1$~pc and $R_{\rm{SC}} = 3$~pc, but setting the radii to smaller values increases the gas density and thus increases \Lx\ (see Section~\ref{sec.Variation.radii}). However, from observations we expect the star clusters to populate a range in mass and size \citep[e.g.][]{Baumgardt.Hilker.2018}, such that density would not scale directly with cluster mass. Since our model does not account for such a spread in cluster sizes, we highlight that \Lx\ may be underestimated for low mass star clusters, whilst overestimated for high mass clusters if the cluster sizes are likewise larger. However, there is only a weak systematic relation between star cluster mass and radius \citep[see e.g.][]{Larsen2004,Krumholz2019}. 

We construct starburst galaxies out of star clusters according to their SFR. To do so, we add star clusters to the starburst galaxies so that the SFR is fulfilled with ${\rm{SFR}} = m_{\rm{min}} / {\rm{d}}t$ (see Section~\ref{sec.galaxies} and Equation~\eqref{eq.mmin}). Since we only keep the star clusters up to an age of 20~Myr, we thus limit the total mass of the starburst galaxy and likewise the maximum mass of the most massive star cluster that can be present (see Equation~\eqref{eq.galmax}). For example, a starburst galaxy with ${\rm{SFR}} = 1$~\Msunyr\ (${\rm{SFR}} = 0.1$~\Msunyr) can only contain star clusters up to $m_{\rm{clus}} \sim 10^7$~$M_{\odot}$ ($m_{\rm{clus}} \sim 10^6$~$M_{\odot}$). Thus the mass range of star clusters limits the range of SFR we can study with our model.

We construct starburst galaxies in a way that the SFR is constant, averaged over time scales of $\leq 20$~Myr. In observations, the formation of stars can take place in several episodes, and thus, the SFR varies over time. When estimated observationally, the SFR is then an average over time spans between a few (from H$\alpha$ measurements) and several hundred (from FUV measurements) Myr. For comparing our models with observations, we ideally need to compare to recent SFRs, since the young massive star clusters are those that dominate the X-ray emission of the starburst galaxy.

\subsection{Interpretation of Green Pea galaxies}

The motivation of this study was to test whether the excess X-ray emission in Green Pea galaxies studied by \citet{Svoboda.etal.2019} can be explained by the emission from hot gas in star clusters. According to our models, the stellar winds do not significantly contribute to the X-ray emission with Green Pea-like metallicities, and are hence unlikely to explain the X-ray excess observed in these sources. We found the X-ray emission to increase when considering larger cluster masses, smaller core and cutoff radii of the star clusters, and a flatter power law index $\beta$ of the cluster mass function. In general, young massive star clusters dominate the X-ray spectrum of the starburst galaxy in our model. To obtain X-ray luminosities of $L_{\rm{X}} \sim 10^{41}$ -- $10^{42}$~\ergs\ as found for the Green Pea galaxies, the starburst galaxy could host massive star clusters. The most massive star clusters our model calculates have $m_{\rm{clus}} \sim 10^8$~$M_{\odot}$, which result in an X-ray luminosity of $L_{\rm{X}} \sim 10^{40}$ -- $10^{41}$~\ergs\ depending on the metallicity (see Sections~\ref{sec.Xray_Starcluster.LscLx} and \ref{sec.Xray_Starcluster.grid}). The Green Pea galaxies have SFRs in the range of \mbox{$\sim 40$ -- $60$~\Msunyr}, which in our model would correspond to $m_{\rm{gal, max}} \sim 10^9$~$M_{\odot}$ (see Equation~\eqref{eq.galmax}). Thus if the complete mass $m_{\rm{gal, max}}$ of a Green Pea galaxy is enclosed in massive star clusters of $10^8$~$M_{\odot}$, we could reproduce the measured luminosity. However, we find it more likely that there are other sources of X-ray emission in the X-ray bright Green Pea galaxies discovered by \citeauthor{Svoboda.etal.2019}. 

In low metallicity environments such as Green Pea galaxies, it is expected that massive stars lose less mass during their lives and thus form heavier black holes than in high-metallicity environments. Therefore, High-Mass X-ray Binaries could be more common in low metallicity environments \citep{Schaerer.etal.2019}, and are thus another possible source for X-ray emission that can be tested in future work.

\section{Summary}
\label{sec.summary}

We calculated the soft X-ray emission in the energy band \mbox{0.5 -- 2.0~keV} from colliding stellar winds in star clusters for a grid of cluster masses and ages, assuming sub-solar metallicities ($0.02$ and $0.4$~$Z_{\odot}$) and including supernova explosions of massive stars. We use these star clusters to construct starburst galaxies with a given SFR defined via the mass function of the star clusters, and to estimate the X-ray emission in the starburst galaxies from the hot gas in star clusters. How we construct the starburst galaxies is based on various assumptions such as the range of cluster masses and the cluster mass function power law index $\beta$. For $Z = 0.02$~$Z_{\odot}$, we find the X-ray emission of our model to underpredict the X-ray luminosities observed in star-forming galaxies \citep[see e.g.][]{Mineo.etal.2012.II}. For $Z = 0.4$~$Z_{\odot}$, we are in good agreement with \citet{Mineo.etal.2012.II} once we include star clusters up to $10^8$~$M_{\odot}$, or choose a flatter cluster mass function index of $\beta \gtrsim -1.5$. We then varied model parameters to study the general behaviour of the X-ray emission under different model assumptions. 
\begin{itemize}[noitemsep]
	\item[\textit{(i)}] We find the X-ray emission to be higher in star clusters and starburst galaxies with larger metallicities. Stars at higher metallicity have stronger mass loss, and therefore the gas density and ultimately the X-ray emission in these clusters is larger.
    \item[\textit{(ii)}] The X-ray emission of the starburst galaxies is dominated by the young massive star clusters, and varies with the assumed cluster mass range. Increasing the range of considered cluster masses towards higher values yields increased X-ray luminosity, as well as the X-ray luminosity fluctuations.
    \item[\textit{(iii)}] When excluding low mass clusters in the considered mass range, the mass of the starburst galaxies is distributed among clusters with higher masses, forcing the X-ray emission of the starburst galaxies higher, and reducing the fluctuations.
    \item[\textit{(iv)}] The power law index $\beta$ of the cluster mass function influences the distribution of the star clusters, and therefore the estimated X-ray emission. Varying the cluster mass function $\beta$ from the fiducial value of $-2$ to a flatter index ($\gtrsim -1.5$) allows a larger number of high mass clusters to be present in the constructed starburst galaxies and thus the X-ray luminosity increases.
    \item[\textit{(v)}] Assuming smaller cluster radii would increase the gas density and cause more X-ray emission. 
    \item[\textit{(vi)}] Including chemically-homogeneous evolving stars also leads to an increased X-ray luminosity, though it is not sufficient to close the gap between our estimated luminosities and those observed for e.g., the Green Pea galaxies.
\end{itemize}
Interpreting these results in the context of the enhanced soft X-ray emission measured in the Green Pea galaxies by \citet{Svoboda.etal.2019}, we conclude that the wind material does not significantly contribute to the X-ray emission. Thus, other sources of X-ray emission must be tested in combination with deeper multi-wavelength observations to disentangle the elusive emission mechanism(s) responsible for such bright X-ray emission in Green Pea galaxies. Possible remaining mechanisms for the Green Pea X-ray excess include: unusually high quantities of High- and/or Low-Mass X-ray Binaries, large quantities of Ultra-Luminous X-ray sources, modified initial mass functions, Intermediate-Mass Black Holes or AGN.


\acknowledgments
We thank the referee for constructive comments which helped improve the paper. A.F. thanks Angela Adamo and Nathalie Webb for fruitful discussions about star clusters and Green Pea galaxies. This work has been supported by the institutional  project  RVO:67985815. AF \& RW acknowledge financial support from the Czech Science Foundation project No. 19-15008S. S.M.G acknowledges support from CONACYT-M\'exico research grant A1-S-28458 and C\'atedra n.482. P.B. \& J.S. acknowledge financial support from the Czech Science Foundation project No. 22-22643S. D.Sz. has been supported by the Alexander von Humboldt Foundation and was funded in part by the National Science Center (NCN), Poland under grant number OPUS 2021/41/B/ST9/00757. For the purpose of Open Access, the author has applied a CC-BY public copyright license to any Author Accepted Manuscript (AAM) version arising from this submission. The authors thankfully acknowledge the computer resources, technical expertise and support provided by the Laboratorio Nacional de Superc\'omputo del Sureste de M\'exico, CONACYT member of the network of national laboratories.

\bibliography{bibliography.bib}{}
\bibliographystyle{aasjournal}

\end{document}